\documentclass[twocolumn]{aastex631}
\usepackage{CJK}
\usepackage{amsmath}
\usepackage{amssymb}
\usepackage{natbib}
\usepackage{rotating}
\defcitealias{Theissen-2022}{T22}
\defcitealias{Kim-2019}{K19}
\graphicspath{{./}{figures/}}

\begin{document}
\begin{CJK*}{UTF8}{gbsn}
\title{The 3D Kinematics of the Orion Nebula Cluster II: Mass-dependent Kinematics of the Inner Cluster}



\author[0000-0002-2612-2933]{Lingfeng Wei (魏凌枫)}
\affiliation{Department of Physics, University of California, San Diego, La Jolla, CA 92093, USA}
\affiliation{Center for Astrophysics and Space Sciences, University of California, San Diego, La Jolla, CA 92093, USA}

\author[0000-0002-9807-5435]{Christopher A. Theissen}
\affiliation{Department of Astronomy and Astrophysics, University of California, San Diego, La Jolla, CA 92093, USA}
\affiliation{Center for Astrophysics and Space Sciences, University of California, San Diego, La Jolla, CA 92093, USA}

\author[0000-0002-9936-6285]{Quinn M. Konopacky}
\affiliation{Department of Astronomy and Astrophysics, University of California, San Diego, La Jolla, CA 92093, USA}
\affiliation{Center for Astrophysics and Space Sciences, University of California, San Diego, La Jolla, CA 92093, USA}

\author[0000-0001-9611-0009]{Jessica R. Lu}
\affiliation{Department of Astronomy, University of California, Berkeley, Berkeley, CA 94720-3411, USA}

\author[0000-0002-5370-7494]{Chih-Chun Hsu}
\affiliation{Center for Interdisciplinary Exploration and Research in Astrophysics (CIERA), Northwestern University, Evanston, IL 60201, USA}

\author[0000-0002-6658-5908]{Dongwon Kim}
\affil{Riiid, 10F, Parnas Tower, 521, Teheran-ro, Gangnam-gu, Seoul 06164, Republic of Korea}

\begin{abstract}
We present the kinematic analysis of $246$ stars within $4\arcmin$ from the center of Orion Nebula Cluster (ONC), the closest massive star cluster with active star formation across the full mass range, which provides valuable insights into the formation and evolution of star clusters on an individual-star basis. High-precision radial velocities and surface temperatures are retrieved from spectra acquired by the NIRSPEC instrument used with adaptive optics (NIRSPAO) on the Keck II 10-m telescope. A three-dimensional kinematic map is then constructed by combining with the proper motions previously measured by the Hubble Space Telescope (HST) ACS/WFPC2/WFC3IR and Keck II NIRC2. The measured root-mean-squared velocity dispersion is $2.26\pm0.08~\mathrm{km}\,\mathrm{s}^{-1}$, significantly higher than the virial equilibrium's requirement of $1.73~\mathrm{km}\,\mathrm{s}^{-1}$, suggesting that the ONC core is supervirial, consistent with previous findings. Energy equipartition is not detected in the cluster. Most notably, the velocity of each star relative to its neighbors is found to be negatively correlated with stellar mass.  Low-mass stars moving faster than their surrounding stars in a supervirial cluster suggest that the initial masses of forming stars may be related to their initial kinematic states. Additionally, a clockwise rotation preference is detected. A weak sign of inverse mass segregation is also identified among stars, excluding the Trapezium stars, though it could be a sample bias. Finally, this study reports the discovery of four new candidate spectroscopic binary systems.

\end{abstract}



\section{Introduction} \label{sec:intro}
Star clusters are the primary sites for a multitude of star formation processes observed throughout the universe \citep[][]{Lada-2003, Gutermuth-2009}. Studying the formation and evolution of star clusters is therefore of crucial importance to constrain star formation theories. The kinematics of star clusters provide valuable insights into the processes of their formation and evolution. Despite contemporary observational efforts, many of the details regarding formation processes in star clusters remain to be unveiled \citep[e.g.,][]{Krumholz-2014}. A particular challenge has been generating models that successfully explain the formation of low-mass stars ($M < 0.3\,M_\Sun$). Initial models of the competitive accretion process naturally explained the formation of low-mass stars by invoking a natal cluster with similarly-sized cores in which some protostars were cut off from the reservoir of material through violent dynamical interactions \citep[e.g.,][]{Bate-2003}. However, competitive accretion models have difficulties in explaining the existence of protoplanetary disks and wide binary systems amongst the lowest mass stars \citep[e.g.,][]{Burgasser-2007}. More recently, simulations in which low-mass stars form along dense filaments of gas infalling into the forming cluster have more successfully predicted the observed multiplicity and disk properties \citep[e.g.,][]{Bonnell-2008, Kainulainen-2017}. The low-mass stars formed via fragmentation within filaments have high velocities that prevent them from accreting additional material from the environment. The signature of this formation process can potentially be observed in very young, non-relaxed clusters as a negative correlation between velocity and mass.  Therefore, observations that probe the kinematics of stars in young clusters have the potential to shed new light on the origin of low-mass stars.

The ONC is an optimal target for the study of the formation and evolution of star clusters, as it is the nearest ($389\pm3$~pc; \citealt{Kounkel-2018}) active massive stellar nursery. At about $2$~Myr \citep[][]{Hillenbrand-1997, Reggiani-2011}, the youth and proximity of the ONC make it ideal for studying the early formation process of a cluster via kinematic measurements, such as radial velocities and proper motions. 

Despite its proximity to us, kinematic observations are plagued by the nebulosity and crowding in the region, especially towards the center of the ONC, where the Trapezium, a collection of the brightest stars at the heart of the ONC, lies. \citet{Furesz-2008} and \citet{Tobin-2009} conducted a large-scale radial velocity (RV) survey of $1215$ and $1613$ stars in the ONC, respectively, using the multi-fiber echelle spectrograph at the $6.5$-m MMT and Magellan telescopes. \citet{Kounkel-2016} presented a reanalysis of the data in \citet{Tobin-2009} as well as more recent supplementary observations. The Apache Point Observatory Galactic Evolution Experiment (APOGEE; \citealt{Majewski-2017}) spectrograph on the $2.5$-m Sloan Digital Sky Survey (SDSS; \citealt{York-2000}) telescope has also acquired near-infrared high-resolution spectroscopic data towards the broader Orion Complex region \citep[][]{DaRio-2016, DaRio-2017, Kounkel-2018}. However, observations mentioned above have limited coverage near the Trapezium due to its dense, crowded, and highly embedded nature. Most recently, the \textit{Gaia} Data Release 3 (DR3; \citealt{Gaia-2016, GaiaDR3-2023j}) provides more complete coverage in the area with more precise astrometric solutions. However, \textit{Gaia} DR3 lacks the spectroscopic data necessary to infer stellar parameters, such as effective temperature and surface gravity, and as an optical mission, is also plagued by nebulosity. 

\citet[hereafter T22]{Theissen-2022} present a kinematic analysis of $56$ sources within $4\arcmin$ of the ONC center observed by Keck II NIRSPAO and $172$ sources observed by APOGEE. The study concludes that the central region of the ONC is not fully virialized by measuring its intrinsic velocity dispersion (IVD). Moreover, the radial IVD is found to be higher than the tangential component as measured by proper motions from the Hubble Space Telescope (HST) + Keck \citep[][hereafter K19]{Kim-2019}. The work presented here expands the sample size of the sources observed by NIRSPAO and presents further kinematic analysis of the region.

In spite of the observations and studies on the ONC over several decades, many questions remain unanswered regarding the kinematic characteristics and the formation process of this young and dense cluster. For instance, the current virial state of the ONC is unclear. \citet{Hillenbrand-1998} suggests that some portion of the ONC is already bounded, and the cluster will eventually become bound. Velocity dispersion measurements seem to confirm that the ONC is moderately supervirial \citep[e.g.,][]{DaRio-2014, DaRio-2017, Kim-2019, Theissen-2022}. On the other hand, the most recent study on the ONC advocates that the ONC is likely bound, as previous measurements of the virial parameter may be inflated due to ejections in unstable N-body interactions \cite{Kounkel-2022}. N-body simulations can rule out neither possibility \citep[][]{Kroupa-2000, Scally-2005}. Another fundamental question is the formation mechanism of the cluster as a whole. \citet{Kounkel-2022} uses the data from \textit{Gaia} DR3 and identifies a stellar age gradient as a function of their distance from us, implying that the star formation front is propagating into the star cluster, possibly triggered by the shockwave from a supernova in the past. Evidence suggests that the ONC could form in the oscillating integral-shaped filament (ISF), a filamentary gas structure associated with the ONC \citep{Bally-1987}, and recently ejected from the ISF \citep{Stutz-2016, Stutz-2018, Matus-2023}. Such a mechanism of producing protostars is referred to asthe  slingshot scenario.  Moreover, a net infall of young stars along the ISF towards the center is discovered \citep[][]{Kounkel-2022}, consistent with the gravitational fragmentation star formation mechanism \citep[][]{Bonnell-2008}.

\begin{figure*}[ht!]
    \centering
    \includegraphics[width=\columnwidth]{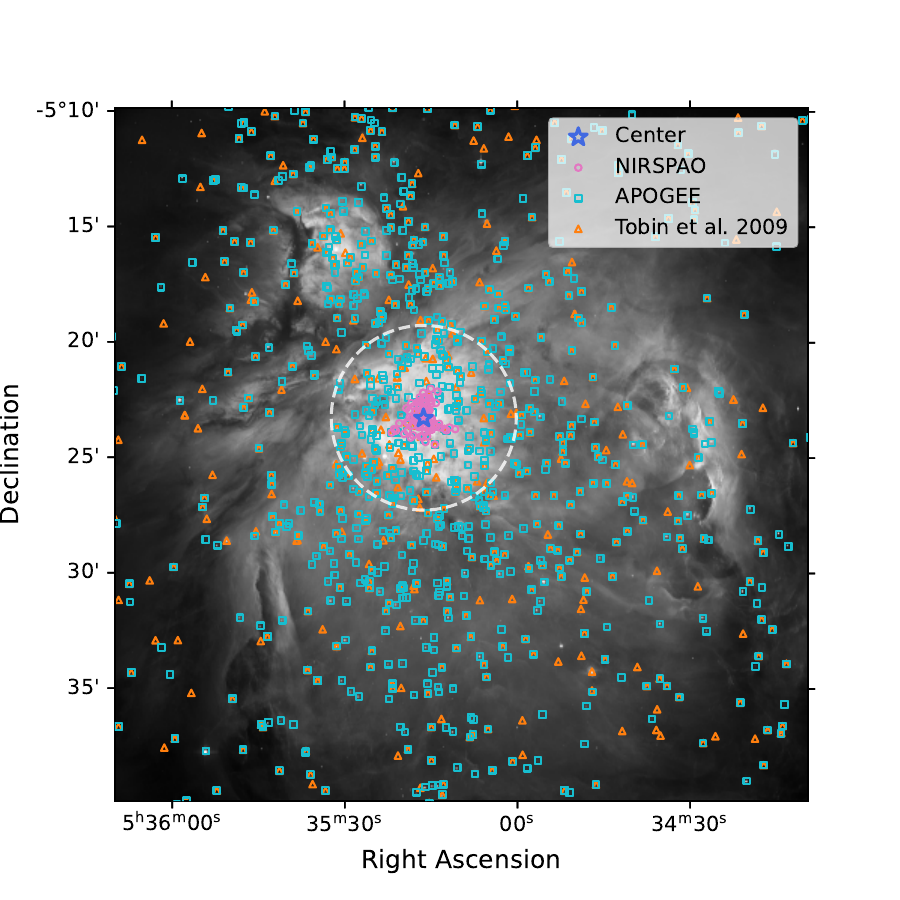}
    \includegraphics[width=\columnwidth]{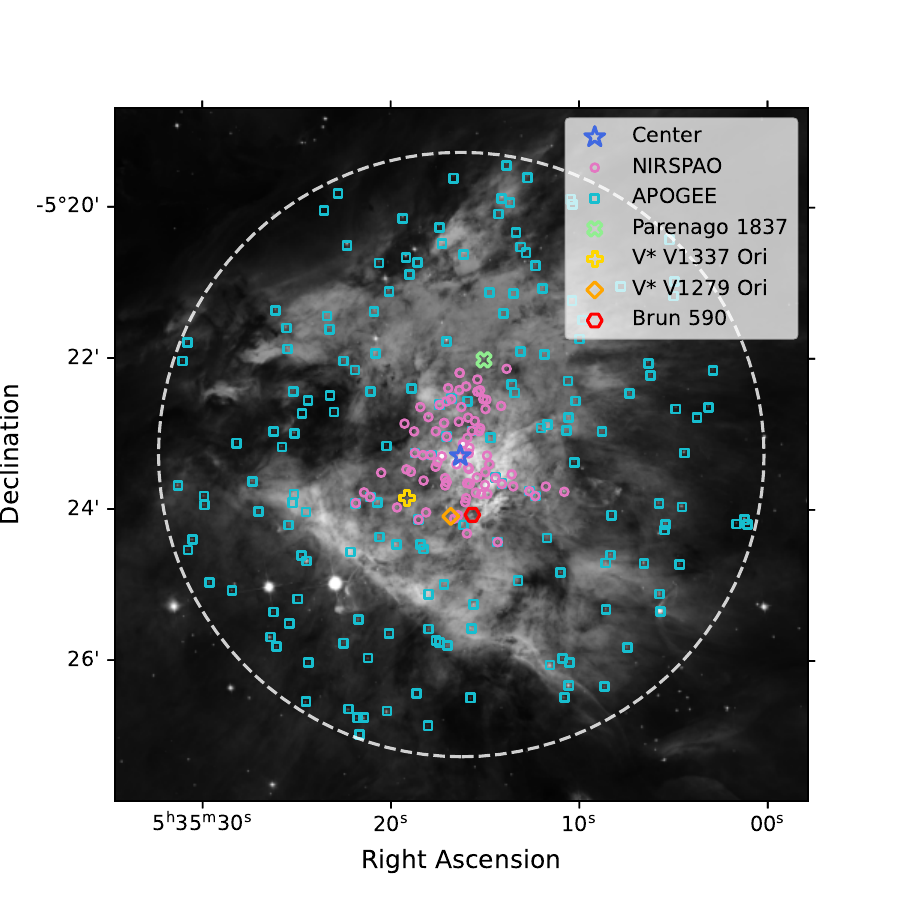}
    \caption{Distribution of sources observed previously and in this study on the background of HST ACS R-band image of the ONC \citep{Robberto-2013}.  \textit{Left:} A wide view of the central $30\arcmin\times30\arcmin$ of the cluster. The $240$ sources observed by NIRSPEC and APOGEE within the circle are selected for analysis in this work. The sources observed by NIRSPAO are marked with magenta circles. Sources observed by APOGEE and by the RV survey in \citet{Tobin-2009} are marked in cyan boxes and amber triangles, respectively. The dashed white circle indicates the $4\arcmin$ radius threshold.  \textit{Right:} A detailed view within the central $4\arcmin$ radius, highlighting the sources considered in this study. The cluster center is labeled as the blue star. Four candidate spectroscopic binary systems identified in this work, namely Parenago 1837, V* V1337 Ori, V* 1279 Ori, and Brun 590, are marked as green plus, yellow cross, amber diamond, and red hexagon, respectively. }
    \label{fig:skymap}
\end{figure*}

In this work, we use the W. M. Keck Observatory to acquire near-infrared (NIR) high-resolution spectra of $91$ sources at the central ONC region, $24$ of which are newly observed after \citetalias{Theissen-2022}. Combined with an updated analysis of $172$ stars observed by APOGEE, astrometric measurements from \textit{Gaia} DR3, and proper motions from the HST + Keck \citepalias{Kim-2019}, a more thorough analysis of the ONC is made possible in this work, pushing the boundary of our understanding of star cluster formation and evolution. 

In Section~\ref{sec:observation}, we introduce the new data from the Keck Observatory and the data reduction processes. Section~\ref{sec:results} presents the results of three-dimensional kinematics, including radial velocity modeled from the spectrum and proper motions measured previously, and the derivation of stellar masses. In Section~\ref{sec:analysis}, we analyze the mass-dependent kinematics of the ONC core, including the virial state, velocity-mass relation, effective temperature offset between NIRSPAO and APOGEE, and preferred proper motion direction. In Section~\ref{sec:binary}, we report the identification of candidate single-lined spectroscopic binaries (SB1): Parenago 1837, V* V1337 Ori, V* 1279 Ori, and Brun 590. Moreover, we simulate the effect of binaries on the velocity dispersion in the section. The implications of our results on kinematic structure, star formation in the ONC are discussed in Section~\ref{sec:discussion}. Mass segregation and binarity in the cluster are also explored in the same section. Lastly, Section~\ref{sec:conclusion} gives a summary of this study and an outlook of our future observation and research plans.

\section{Observation and Data Reduction}
\label{sec:observation}

\subsection{Sample Selection}
\label{subsec:sample}

Our sample consists of a total of $246$ sources with 2D positions and 3D velocities. $91$ of them are observed by NIRSPAO on Keck II within $2\arcmin$ of the central Trapezium, and $172$ sources are re-analyzed APOGEE sources within $4\arcmin$ of the ONC core, with $17$ sources overlapping.

The NIRSPAO sources were chosen based on their inclusion in the proper motion catalog presented in \citetalias{Kim-2019}, as our goal is to measure the three-dimensional motion of the stars in our sample. The brightness of the sample ranges from $7.4$ to $12.9$ magnitude in K band.  The total size of the sample targeted with new observations of $91$ was primarily driven by observing time constraints from an initial selection of $100$ targets in the central ONC.  The sources were selected to span a range of magnitudes, which roughly corresponds to a range of temperatures and masses.  Since our goal is to assess the kinematic behavior of the lowest mass sources, attention was paid to targeting a sufficient number of faint sources to place statistical constraints on their motion.  A sky map of the sources targeted in this study is illustrated in Figure~\ref{fig:skymap}. Sources observed in this study are pinpointed by magenta circles. 

We supplement the sample observed with an updated analysis of $172$ stars observed by APOGEE within $4\arcmin$ of the center of the ONC \citepalias[][]{Theissen-2022}. APOGEE targets are marked in cyan boxes and the RV survey conducted by \citet{Tobin-2009} and reanalyzed by \citet{Kounkel-2016} are represented as amber triangles in Figure ~\ref{fig:skymap}. The dashed white circle indicates the $4\arcmin$-radius from the center. Throughout this work, we adopted the same location used by \citep[][]{DaRio-2014} as the center of mass of the ONC: $\alpha_{J2000}=05^h35^m16.26^s, \delta_{J2000}=-05^\circ23\arcmin16.4\arcsec$. All $240$ sources in our sample are located within the circle. A zoom-in on this region is shown in the right panel of Figure~\ref{fig:skymap}. As can be seen, previous studies do not have extensive coverage of the central region due to its crowded nature and high level of nebulosity. AO-fed observations in the near-infrared greatly aid in the measurement of individual spectra in this region.

\subsection{Observation}
\label{subsec:observation}

To observe our $91$ targets, we utilized NIRSPEC in conjunction with the Keck II facility laser guide star (LGS) adaptive optics (AO) system \citep[][]{McLean-1998, McLean-2000, vanDam-2006, Wizinowich-2006, Martin-2018}. NIRSPEC is a near-infrared echelle spectrograph on Keck II. The observations were conducted between $2015$ to $2022$. The number of sources observed with NIRSPEC with AO (NIRSPAO) increased from $56$ to $91$ compared to the previous study \citepalias[][]{Theissen-2022}, a $\sim63\%$ increase.  Exposures with NIRSPAO utilize the $0.041\times2.26"$ slit in the the NIRSPEC-7 filter and K-new filter before and after the upgrade, covering the wavelength of $1.839$--$2.630$ and $1.907$--$2.554$, respectively. This wavelength regime covers the carbon monoxide (CO) absorption lines around $2.29$--$2.40~\mu m$, which are present in the spectra of low-mass stars. Moreover, the hydrogen Brackett-$\gamma$ line at $2.166~\mu m$ and the Si, Fe, and Ti lines at $2.18$--$2.19\mu m$ are also within the wavelength range, which helps in inferring the stellar parameters of higher-mass stars. The spectrograph splits the incoming starlight into multiple rows so as to fit in the square-shaped detector, and each row is referred to as an order. In this work, the wavelength coverage of each order in our setup of the detector offset is $2.044$--$2.075~\mu m$ for order $37$, $2.100$--$2.133~\mu m$ for order $36$, $2.162$--$2.193~\mu m$ for order $35$, $2.224$--$2.256~\mu m$ for order $34$, $2.291$--$2.325~\mu m$ for order $33$, and $2.362$--$2.382~\mu m$ for order $32$. The CO lines fall within the range of orders $32-33$, while the Si, Ti, and Fe lines are situated in order $35$ for sources with higher effective temperatures. Therefore, we primarily use orders $32$, $33$, and $35$ to sample the stellar parameters.  The resolution of the spectra is $\mathrm{R}\sim25000$ for data collected before 2019 and $\mathrm{R}\sim35,000$ on or after 2019 as a result of the upgrade on Keck.

While some targets were bright enough to serve as a natural guide star ($R\lesssim15$), the extinction in this region means that most of our sources required LGS.  There are sufficient sources in the region to supply the needed R$\sim$18 tip/tilt guide star requirement within 1' of the target \citep{Wizinowich-2006}.  For the majority of observations, the target was acquired with $\mathrm{PA}=0^\circ$.  However, in some cases there were two sources close enough together to position on the slit simultaneously.  In those instances, we rotated to an appropriate PA to align both stars to the fall on the slit.  

For the majority of the targets, we take four exposures by placing the sources in the slit in an upper-lower-lower-upper sequence, or ABBA dither pattern. In a few cases, the number of frames differs from $4$ due to either loss of target or interruption of observation. HD$37887$, a star of spectral type B9.5IV/V, is used as the calibration star at a similar airmass for telluric wavelength adjustment either before or after a science object is observed.  A log of all NIRSPAO observations, including the dates and the total time of source, is given in Table~\ref{tab:observation}. 

\startlongtable
\begin{deluxetable}{lccc}
\tablecaption{Log of NIRSPAO Observations}
\label{tab:observation}
\tablecolumns{4}
\tablehead{
\colhead{HC2000 ID} & \colhead{Date of} & \colhead{No. of} & \colhead{Exp. Time}\\[-0.2cm]
\colhead{} & \colhead{Obs. (UT)} & \colhead{Frames\tablenotemark{a}} & \colhead{Per Frame (s)}
}
\startdata
 HC2000 322 & 2015 Dec 23 &             4 &       300 \\
 HC2000 296 & 2015 Dec 23 &             4 &      1200 \\
 HC2000 259 & 2015 Dec 23 &             4 &        90 \\
 HC2000 213 & 2015 Dec 23 &             4 &        60 \\
HC2000 306A & 2015 Dec 24 &             4 &       180 \\
HC2000 306B & 2015 Dec 24 &             4 &       180 \\
HC2000 291A & 2015 Dec 24 &             4 &       600 \\
HC2000 291B & 2015 Dec 24 &             4 &       600 \\
 HC2000 252 & 2015 Dec 24 &             4 &       300 \\
 HC2000 250 & 2015 Dec 24 &             4 &      1200 \\
 HC2000 244 & 2015 Dec 24 &             4 &       600 \\
 HC2000 261 & 2015 Dec 24 &             4 &       900 \\
 HC2000 248 & 2016 Dec 14 &             4 &       600 \\
 HC2000 223 & 2016 Dec 14 &             4 &       300 \\
 HC2000 219 & 2016 Dec 14 &             4 &       600 \\
 HC2000 324 & 2016 Dec 14 &             3 &      1200 \\
 HC2000 295 & 2018 Feb 11 &             4 &       450 \\
 HC2000 313 & 2018 Feb 11 &             4 &       180 \\
 HC2000 332 & 2018 Feb 11 &             4 &       300 \\
 HC2000 331 & 2018 Feb 11 &             4 &       450 \\
 HC2000 337 & 2018 Feb 11 &             4 &        60 \\
 HC2000 375 & 2018 Feb 11 &             4 &       180 \\
 HC2000 388 & 2018 Feb 11 &             4 &       120 \\
 HC2000 425 & 2018 Feb 12 &             4 &        60 \\
 HC2000 713 & 2018 Feb 12 &             4 &        90 \\
 HC2000 408 & 2018 Feb 12 &             4 &       450 \\
 HC2000 410 & 2018 Feb 12 &             4 &       600 \\
 HC2000 436 & 2018 Feb 12 &             4 &        90 \\
 HC2000 442 & 2018 Feb 13 &             4 &       450 \\
HC2000 522A & 2019 Jan 12 &             4 &       450 \\
HC2000 522B & 2019 Jan 12 &             4 &       450 \\
 HC2000 145 & 2019 Jan 12 &             4 &       600 \\
 HC2000 202 & 2019 Jan 12 &             4 &       120 \\
 HC2000 188 & 2019 Jan 12 &             4 &       600 \\
 HC2000 302 & 2019 Jan 13 &             4 &       450 \\
 HC2000 275 & 2019 Jan 13 &             2 &       450 \\
 HC2000 245 & 2019 Jan 13 &             4 &       180 \\
 HC2000 258 & 2019 Jan 13 &             4 &       180 \\
 HC2000 220 & 2019 Jan 13 &             4 &       600 \\
 HC2000 370 & 2019 Jan 16 &             4 &       180 \\
 HC2000 389 & 2019 Jan 16 &             4 &       120 \\
 HC2000 386 & 2019 Jan 16 &             4 &       120 \\
 HC2000 398 & 2019 Jan 16 &             4 &       120 \\
 HC2000 413 & 2019 Jan 16 &             4 &       180 \\
 HC2000 253 & 2019 Jan 16 &             4 &       120 \\
 HC2000 288 & 2019 Jan 17 &             4 &       450 \\
 HC2000 420 & 2019 Jan 17 &             4 &       450 \\
 HC2000 412 & 2019 Jan 17 &             4 &       450 \\
 HC2000 282 & 2019 Jan 17 &             4 &       450 \\
 HC2000 217 & 2019 Jan 17 &             4 &       180 \\
 HC2000 217 & 2020 Jan 18 &             4 &       120 \\
 HC2000 229 & 2020 Jan 18 &             4 &       450 \\
 HC2000 228 & 2020 Jan 19 &             4 &       120 \\
 HC2000 224 & 2020 Jan 19 &             4 &        90 \\
 HC2000 135 & 2020 Jan 19 &             4 &       180 \\
 HC2000 440 & 2020 Jan 20 &             4 &       120 \\
 HC2000 450 & 2020 Jan 20 &             4 &       300 \\
 HC2000 277 & 2020 Jan 20 &             4 &       300 \\
 HC2000 204 & 2020 Jan 20 &             4 &       300 \\
 HC2000 229 & 2020 Jan 20 &             4 &       450 \\
 HC2000 214 & 2020 Jan 20 &             4 &       450 \\
 HC2000 215 & 2020 Jan 21 &             4 &       300 \\
 HC2000 240 & 2020 Jan 21 &             4 &       300 \\
 HC2000 546 & 2020 Jan 21 &             4 &       300 \\
 HC2000 504 & 2020 Jan 21 &             4 &       120 \\
 HC2000 703 & 2020 Jan 21 &             4 &       300 \\
 HC2000 431 & 2020 Jan 21 &             4 &       120 \\
 HC2000 229 & 2020 Jan 21 &             3 &       450 \\
 HC2000 484 & 2021 Feb 01 &             4 &       300 \\
 HC2000 476 & 2021 Feb 01 &             6 &       300 \\
 HC2000 546 & 2021 Oct 20 &             4 &       300 \\
 HC2000 217 & 2021 Oct 20 &             4 &       120 \\
 HC2000 277 & 2021 Oct 20 &             2 &       600 \\
 HC2000 435 & 2021 Oct 20 &             3 &       600 \\
 HC2000 457 & 2022 Jan 18 &             4 &       750 \\
 HC2000 479 & 2022 Jan 18 &             4 &       900 \\
 HC2000 490 & 2022 Jan 18 &             4 &       300 \\
 HC2000 478 & 2022 Jan 18 &             4 &       420 \\
 HC2000 456 & 2022 Jan 18 &             4 &       900 \\
 HC2000 170 & 2022 Jan 18 &             4 &       120 \\
 HC2000 453 & 2022 Jan 19 &             4 &       600 \\
 HC2000 438 & 2022 Jan 19 &             4 &       300 \\
 HC2000 530 & 2022 Jan 19 &             4 &       900 \\
 HC2000 287 & 2022 Jan 19 &             4 &       300 \\
 HC2000 171 & 2022 Jan 19 &             4 &       120 \\
 HC2000 238 & 2022 Jan 20 &             4 &       600 \\
 HC2000 266 & 2022 Jan 20 &             4 &       600 \\
 HC2000 247 & 2022 Jan 20 &             4 &       300 \\
 HC2000 172 & 2022 Jan 20 &             4 &       300 \\
 HC2000 165 & 2022 Jan 20 &             4 &       450 \\
 HC2000 177 & 2022 Jan 20 &             4 &       420 \\
 HC2000 163 & 2022 Jan 20 &             3 &       420 \\
\enddata
\tablenotetext{a}{Number of ABBA frames.}
\end{deluxetable}

\subsection{Data Reduction}
\label{subsec:reduction}
NIRSPEC Data Reduction Pipeline (NSDRP)\footnote{\url{https://github.com/Keck-DataReductionPipelines/NIRSPEC-Data-Reduction-Pipeline}} is a pipeline specifically designed for reducing NIRSPEC spectra and is optimized for point sources. In this work, data reduction is conducted using a modified version of the NSDRP\footnote{\url{https://github.com/ctheissen/NIRSPEC-Data-Reduction-Pipeline}}. The modification includes spatial rectification using the object trace instead of the order edge traces, spectral rectification and wavelength calibration using etalon lamps, cosmic-ray cleaning of flats, and bad-pixel cleaning \citep[see][for details]{Hsu-2021-paper, Hsu-2021-code}

The steps to reduce the data for each source are briefly summarized below.
\begin{enumerate}
    \item Median combines the flat frames to generate a master flat frame, which is used to find order edges.
    \item Run the modified NSDRP pipeline to reduce all the frames after trimming the spectra edges.
    \item Perform initial wavelength calibration for each order using etalon or sky lines of the telluric spectra.
\end{enumerate}

The reduced spectra are then forward modeled for stellar parameters, which will be discussed in Section~\ref{subsec:modeling}.

\subsection{Spectral Forward Modeling}
\label{subsec:modeling}
The reduced spectra are coadded and forward-modeled to derive the stellar parameters. Instead of modeling each individual exposure of the same source as in \citetalias{Theissen-2022}, we coadd the spectra before forward-modeling. Compared to modeling each individual exposure, coaddition helps reduce the white noise, enhancing the signal-to-noise ratio (SNR) of the data, and saving computational resources for spectral forward modeling. The specific steps of coaddition are summarized below. First, the flux of each exposure of the same source is scaled to match the median flux of the frame with the highest signal-to-noise ratio (SNR). It is worth mentioning that scaling does not affect the modeling results. The noise is scaled by the same factor. Next, the fluxes of all the exposures on the same target are averaged, weighted by the inverse square of the corresponding noise on a pixel-wise basis. The noise associated with the coadded spectrum is calculated from the uncertainty propagation equation for weighted averaging. The calculation of the weighted-averaged flux and the noise are illustrated in Equation~\ref{eq:flux} and Equation~\ref{eq:noise}, respectively.

\begin{equation}
    \label{eq:flux}
    f_\mathrm{coadd} = \frac{\sum_{i}\left(f_i/\sigma_i^2\right)}{\sum_{i}\left(1/\sigma_i^2\right)}~,
\end{equation}
\begin{equation}
    \label{eq:noise}
    \sigma_\mathrm{coadd}^2 = \sum_{i}\left(\frac{\partial f_\mathrm{coadd}}{\partial f_i}\cdot\sigma_i\right)^2 = \frac{1}{\sum_{i}\left(1/\sigma_i^2\right)}~,
\end{equation}
where $f_\mathrm{coadd}$ and $\sigma_\mathrm{coadd}$ are the coadded flux and noise. $f_i$ and $\sigma_i$ are the flux and noise of the $i$-th frame for a source.

The coadded spectra are then forward modeled using the Spectral Modeling Analysis and RV Tool \citep[SMART\footnote{\url{https://github.com/chihchunhsu/smart}}, ][]{Hsu-2021-code}. We refer the readers to \citet{Hsu-2021-code, Hsu-2021-paper} and \citetalias{Theissen-2022} for a detailed description of the modeling procedure. The steps of modeling spectra are briefly outlined below.

The first step is obtaining a precise absolute wavelength solution. A quadratic polynomial provided by the NSDRP is adopted as the initial wavelength solution\footnote{\url{https://www2.keck.hawaii.edu/koa/nsdrp/documents/NSDRP_Software_Design.pdf}}. A more precise wavelength solution with a systematic uncertainty of $0.058~\mathrm{km}\,\mathrm{s}^{-1}$ is derived by cross-correlating the spectrum of our A-star calibrator, HD $37887$, and a high-resolution reference telluric spectrum \citep[][]{Moehler-2014} in an iterative approach \citepalias{Theissen-2022}. The coefficients of the polynomial are updated in each iteration by fitting the best wavelength shifts for all cross-correlation windows of 100 pixels. The coadded spectrum is calibrated by the telluric frame in order $32$, $33$, and $35$ with the lowest root-mean-square of the residual for the final wavelength solution.

Next, we use the PHOENIX ACES AGSS COND stellar atmospheric models \citep{Husser-2013} to forward-model the coadded stellar spectrum via the Markov chain Monte Carlo (MCMC) method \citep[][]{Butler-1996, Blake-2007, Blake-2008, Blake-2010, Burgasser-2016} realized by the ensemble sampler \textit{emcee} \citep[][]{Foreman-Mackey-2013}. The flux is modeled by the same function of wavelength as in \citetalias{Theissen-2022}. Note that the surface gravity $\log g$ can hardly be constrained from the spectral modeling within the observed wavelength range and is therefore fixed to be $4$, which is the expected value for young, low-mass stars at the age of the ONC and is consistent with other studies \citep[e.g.,][]{Kounkel-2018}. To see how a different $\log g$ might affect the stellar mass estimation based on modeled effective temperature, we conducted a test on a subset of sources spanning across the temperature and SNR range by vigorously changing $\log g$ to $3.5$ and $4.5$, respectively. The majority of the resulting stellar masses remain within $1\sigma$ of the values under the assumption of $\log g=4$. Only a few sources deviate by $2\sigma$ or more due to a trade-off between surface gravity and temperature with the veiling parameter acting as a tuning knob. Overall, the choice of $\log g=4$ is rationalized as the stellar masses largely remain consistent under a relatively wide range of $\log g$. In addition, metallicity is set to be $0$ based on the average value of the ONC \citep[e.g.,][]{D'Orazi-2009}. 

The free parameters we sample and their corresponding limits and initial distribution are summarized in Table~\ref{tab:prior}. The veiling parameter is defined in the same way as in \citetalias{Theissen-2022}.

Each source is sampled with $100$ walkers and $300$ steps using the \texttt{KDEMove}, discarding the first 200 steps, as the walkers typically converge within the first $100$ steps based on the walker plots. We have also verified the consistency of the results by running the MCMC sampler for $500$ steps, thereby ensuring convergence. A fine-tuning sampling with the same number of walkers, steps, and prior distributions follows after removing the pixels where the residual deviates from its median value by more than three standard deviations of itself. The masking removes the remaining bad pixels and cosmic rays from the spectrum that were not rejected by the NSDRP. The final distribution of the last $100$ steps of the $100$ walkers is considered as the posterior distribution. We take the median of the posterior distribution as the measured value for each parameter, and half the difference between the $16$-th and $84$-th percentile, or the $1$--$\sigma$ range for a normal distribution, as the associated uncertainty. Heliocentric RVs are corrected for barycentric motion using the \textit{astropy}
function \texttt{radial\_velocity\_correction}.

In addition to \texttt{emcee}, we also attempted to use another Bayesian inference tool, \texttt{PyMultiNest} \citep[][]{PyMultiNest}, to sample the posterior distribution of the stellar parameters, adopting the limits in Table~\ref{tab:prior} as the priors. The built-in multimodal nested sampling algorithm is expected to have a better performance in sampling multimodal distributions, which helps with disentangling potential degenerate distributions between the effective temperature and veiling parameter when either one of them is high. However, as we will show in Section~\ref{subsec:3d velocities}, most of the sources have a low veiling parameter, suppressing the degeneracies. Additionally, we noticed a significant increase in the running time as the number of modeled parameters increases. With $9$--$10$ parameters to sample, \textit{emcee} turns out to be the better option in terms of computation efficiency, which is what we eventually adopted, as \textit{emcee} can also sample multimodal distributions unbiasedly.

Reanalyzed results of the APOGEE samples within $4\arcmin$ of the center of the ONC were performed by \citetalias{Theissen-2022}. The identical sampling procedure that is used on the NIRSPEC data is applied to the APOGEE H-band data to model stellar parameters, including effective temperatures, rotational velocities, and RVs for consistency in methodology. The results conform well with the SDSS/APOGEE results.

\begin{deluxetable*}{lcccr}[htb!]
\tablecaption{Spectral Forward-Modeling Free Parameters and Their Bounds} \label{tab:prior}
\tablecolumns{5}
\tablehead{
\colhead{Stellar Parameter} & \colhead{Notation} & \colhead{Limit} & \colhead{Initial Distribution\tablenotemark{a}} & \colhead{Unit} 
}
\startdata
Effective Temparature & $T_\mathrm{eff}$ & $\left(2300, 7000\right)$ & $U\left(2300, 7000\right)$ & K \\
Rotational Velocity & $v\sin{i}$ & $\left(0, 100\right)$ & $U\left(0, 40\right)$ & $\mathrm{km}\,\mathrm{s}^{-1}$ \\
Radial Velocity (RV) & $V_r$ & $\left(-100,100\right)$ & $U\left(-100, 100\right)$ & $\mathrm{km}\,\mathrm{s}^{-1}$ \\
Airmass & $AM$ & $\left(1, 3\right)$ & $U\left(1, 3\right)$ & ---\\
Precip. Water Vapor & $PWV$ & $\left(0.5, 10\right)$ & $U\left(0.5, 10\right)$ & mm \\
Line Spread Function & $\Delta \nu_\mathrm{inst}$ & $\left(1, 20\right)$ & $U\left(1, 10\right)$ & $\mathrm{km}\,\mathrm{s}^{-1}$ \\
Veiling Param.\tablenotemark{b} & $C_\mathrm{veil}$ & $\left(0, 10^{20}\right)$ & $U\left(0, 10^5\right)$ & ---\\
Noise Factor & $C_\mathrm{noise}$ & $\left(1, 50\right)$ & $U\left(1, 5\right)$ & ---\\
Wavelength Offset & $C_\lambda$ & $\left(-1, 1\right)$ & $U\left(-0.2, 0.2\right)$ & \AA \\
\enddata
\tablenotetext{a}{$U\left(a, b\right)$ denotes uniform distribution between $a$ and $b$.}
\tablenotetext{b}{The veiling parameter is defined in the same way as in \citetalias{Theissen-2022}.}
\end{deluxetable*}
For most of the sources, we model orders $32$ and $33$ simultaneously for the stellar parameters. This is where the CO lines reside for low-mass stars. However, this procedure fails for a few high-temperature sources because the spectra in orders $32$ and $33$ are mostly flat without any notable features. Therefore, we modeled Brackett-$\gamma$, silicon, titanium, and iron lines in order $35$ for two high-temperature sources, HC$2000~291$A and HC$2000~337$.

\begin{figure*}[htb!]
    \centering
    \includegraphics[width=\textwidth]{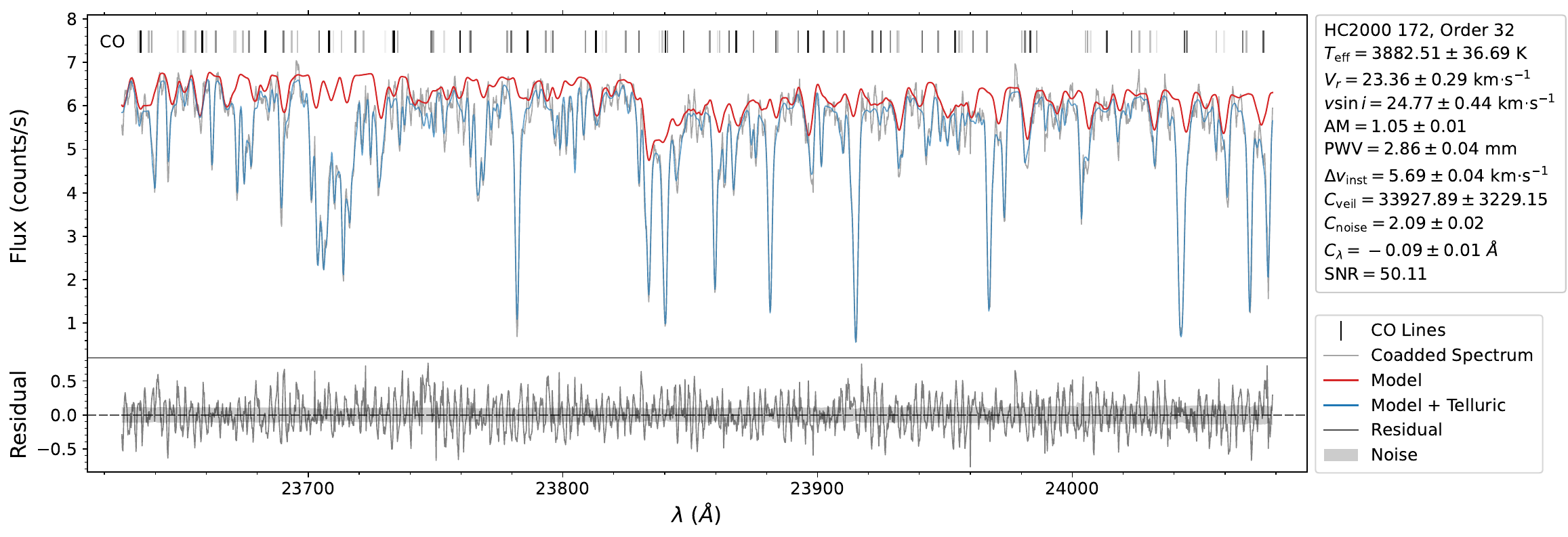}
    \includegraphics[width=\textwidth]{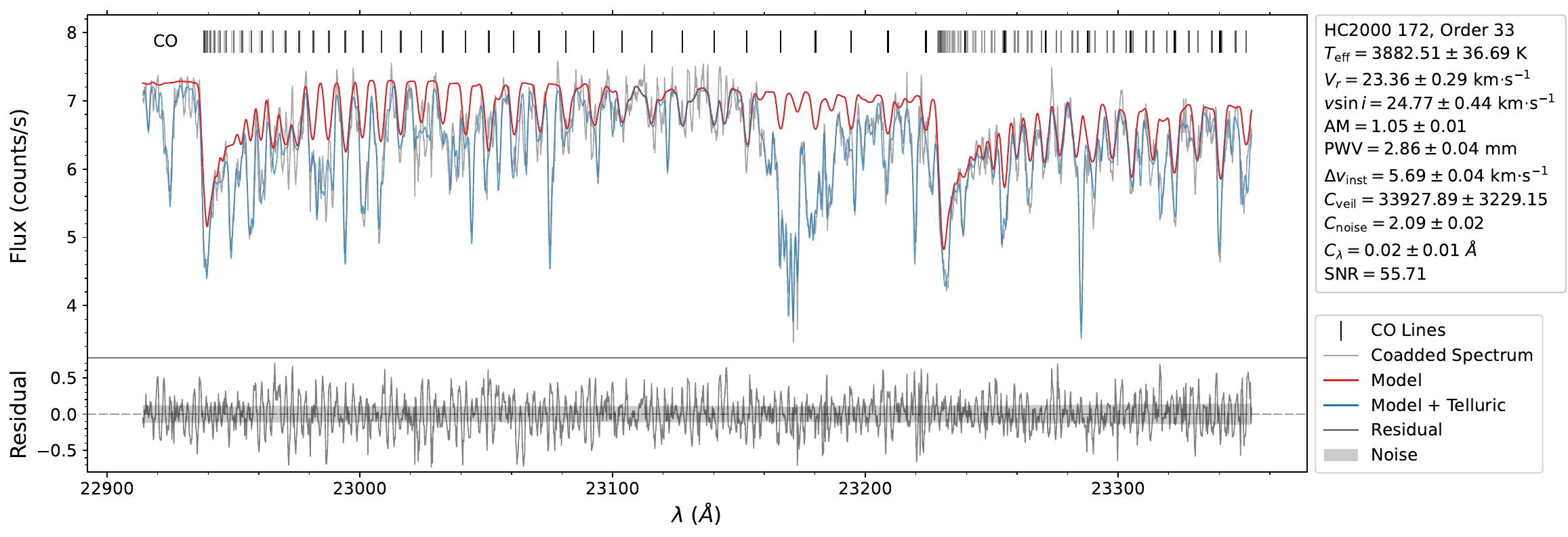}
    \includegraphics[width=\textwidth]{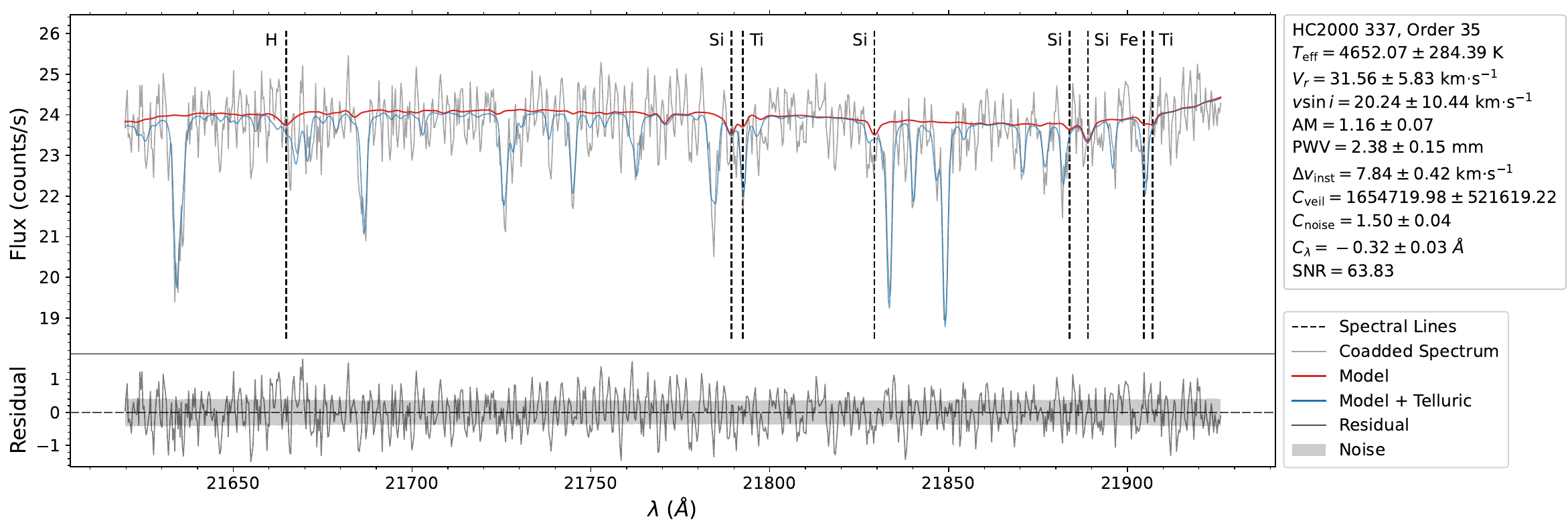}
    \caption{Examples of observed and modeled spectra in different orders from NIRSPAO along with atomic and molecular lines. \textit{Top:} HC2000 172 in order $32$; \textit{Middle:} HC2000 172 in order $33$; \textit{Bottom:} HC2000 337 in order $35$. The upper panel in each figure shows the observed spectrum and the model. The vertical lines denote the locations of atomic and molecular spectral lines. The gray line is the normalized observed spectrum flux. The model with and without telluric features are represented as blue and red lines, respectively. The lower panel shows the residual as the black line and the noise as the shaded area. The modeled parameters of each order are labeled beside the corresponding panels. \label{fig:spectrum}}
\end{figure*}

From top to bottom, Figure~\ref{fig:spectrum} shows an example of the spectrum and model in order $32$ and order $33$ for HC2000 172, and order $35$ for HC2000 337, respectively. The top panel in each figure shows the observed spectrum as the gray line, the model as the red line, and the model multiplied by the telluric model as the blue line. CO lines in order $32$ and $33$ are marked on top of the spectra, with their transparency indicating the corresponding lab intensity according to the HITRAN database \footnote{\url{https://www.spectralcalc.com/spectral_browser/db_data.php}}. We implemented a cut of intensity larger than $10^{-25}~\mathrm{cm}\,\mathrm{mol}^{-1}$ for order $32$ for visualization purposes. The intensity is then normalized from $0.05$ to $0.95$ as transparency. Bracket-$\gamma$, silicon, titanium, and iron lines are indicated as vertical dashed lines in the case where we modeled order $35$ in the bottom figure. The bottom panel in each figure shows the residual as the black line and the noise as the shaded area. Typically, the residual is less than $5\%$ of the median flux. According to the modeling results, HC2000 172 is a source with $T_\mathrm{eff}=3882.5\pm36.7~\mathrm{K}$, $\mathrm{RV}=39.25\pm0.29~\mathrm{km}\,\mathrm{s}^{-1}$, and $v\sin{i}=23.36\pm0.44~\mathrm{km}\,\mathrm{s}^{-1}$. The stellar parameters for HC2000 337 are $T_\mathrm{eff}=4652.1\pm284.4~\mathrm{K}$, $v\sin{i}=20.24\pm10.44~\mathrm{km}\, \mathrm{s}^{-1}$, and $\mathrm{RV}=31.56\pm5.83~\mathrm{km}\, \mathrm{s}^{-1}$. The uncertainty of order $35$ fitting results is still significant, even though it is more than $3$ times better than the $32$ and $33$ fits. Caution should be taken when using modeled parameters in order $35$.

Note that the results might be impacted by the fringing in the spectra, which is the primary contributor to the residuals of spectral modeling (\citealt{Hsu-2021-paper}; \citetalias{Theissen-2022}). But it is unlikely to change most of our results.

\section{Results}
\label{sec:results}

\subsection{Three-Dimensional Velocities}
\label{subsec:3d velocities}

\begin{figure}[htb!]
    \centering
    \includegraphics[width=\columnwidth]{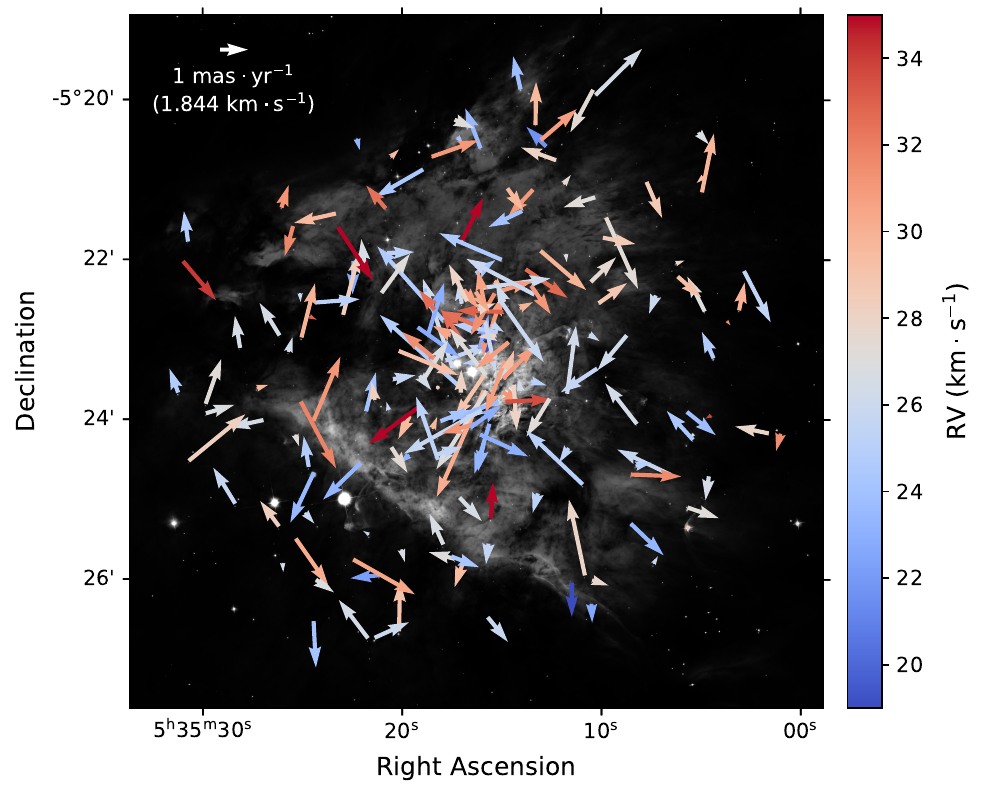}
    \caption{Three-dimensional kinematics of the central $4\arcmin$ of the ONC. The proper motions are denoted by the direction and length of the arrows, and the radial velocities are illustrated by the color. A $1~\mathrm{mas}\,\mathrm{yr}^{-1}$ key to the quiver plot is shown on the top left, or about $1.844~\mathrm{km}\,\mathrm{s}^{-1}$ assuming a distance of $389$~pc.}
    \label{fig:3d kinematics}
\end{figure}

\begin{figure*}
    \centering
    \includegraphics[width=\columnwidth]{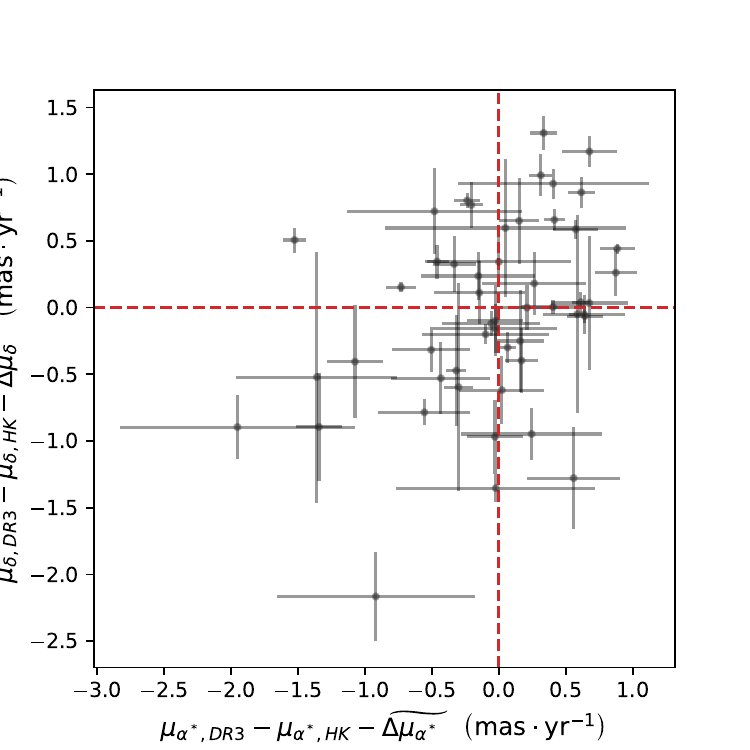}
    \includegraphics[width=\columnwidth]{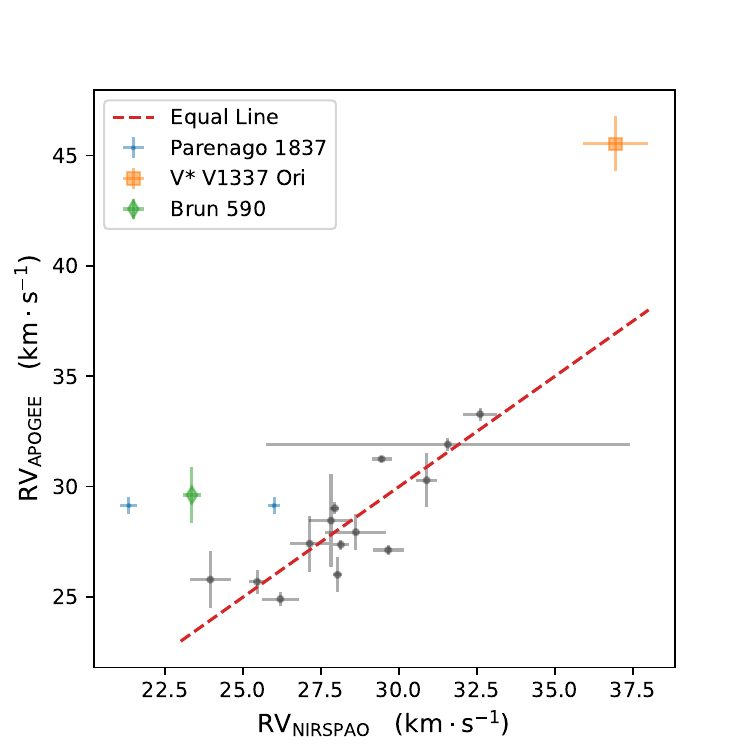}
    \caption{Kinematics comparison with previous measurements. \textit{Left:} Proper motion comparison between \textit{Gaia} DR3 in the absolute frame and HST + Keck in the rest frame. Proper motion in the right ascension and declination are denoted as $\mu_{\alpha^*}$ and $\mu_\delta$, and the measurements from \textit{Gaia} DR3 and \textit{HST} + Keck are denoted as DR3 and HK in the subscript, respectively. In the scope of this work, we transformed the \textit{Gaia} DR3 proper motions into the same reference frame as \textit{HST} + Keck measurements by offsetting the former ones by the average difference between them in both directions. The offsets in right ascension and declination are $\left(\widetilde{\Delta\mu_{\alpha^*}},~\widetilde{\Delta\mu_{\delta}}\right) = \left(1.60,~0.08\right) \mathrm{mas}\,\mathrm{yr}^{-1}$. The red dashed line indicates where the two measurements are equal in RA and DEC directions. \textit{Right:} RV comparison between sources measured with NIRSPAO and APOGEE. The dashed red line indicates the equal line. $3$ of the $4$ candidate binary systems, Parenago 1837, V* V1337 Ori, and Brun 590, are marked in blue dots, amber squares, and green diamonds with errorbars, respectively.}
    \label{fig:kinematics comparison}
\end{figure*}

With the stellar parameters derived from spectral modeling and previous measurements of proper motions and parallaxes, a 3D mass-dependent kinematic map of the ONC core can be constructed. 

When there are multiple epochs of observations for the same object in the NIRSPAO sources, we take the average of the stellar parameters weighted by the inverse of the square of the associated uncertainties. In the case where there is a match between NIRSPEC and APOGEE sources within $1\arcsec$, the radial velocity derived from the NIRSPAO observation is prioritized over the value from APOGEE since NIRSPAO has a higher resolution($25000$ or $35000$ versus $22500$) and to keep consistency with previous works \citepalias[e.g.,][]{Theissen-2022}.

The NIRSPAO and APOGEE sources are then cross-matched with both the proper motion catalog measured by the \textit{HST} + Keck \citepalias[][]{Kim-2019}, and \textit{Gaia} DR3 \citep[][]{Gaia-2016, GaiaDR3-2023j} for parallax within a separation of $1\arcsec$ to construct a 3D kinematic map of the ONC core. We retrieved the \textit{Gaia} DR3 data using \texttt{astroquery} \citep[][]{astroquery}. Due to the low quality of astrometric measurements with Gaia induced by the nebulosity and extinction in the ONC region, we adopted the same generous quality cut of \texttt{astrometric\_gof\_al} $ < 16$ and \texttt{photometric\_mean\_g\_mag} $ < 16$ for \textit{Gaia} DR3 sources in the region selected for cross matching as in \citetalias{Kim-2019}. The G magnitude of the cross-matched sources ranges from $7.8$--$16.0$. The proper motion measurements of \textit{HST} + Keck are prioritized over \textit{Gaia} DR3 for the same concern of astrometric measurement quality in the latter.

To ensure data quality, we applied several constraints on our sample. First, the NIRSPAO and reanalyzed APOGEE sources with a modeled RV uncertainty of no greater than $5~\mathrm{km}\,\mathrm{s}^{-1}$ are selected. $237$ sources out of $246$ remain after the RV constraint is applied. Furthermore, considering our estimated distance from the ONC of $389\pm3$~pc \citep[][]{Kounkel-2018} and the limited accuracy of Gaia astrometric solutions in the region, a generous distance constraint on our sources is imposed with a minimum distance of $300$ pc, a maximum of $500$ pc, and a minimum parallax over error \texttt{parallax\_over\_error} of $5$. $2$ additional source is filtered out in this step, leaving $235$ sources in total. For the remaining sources, we assume the same distance of $389\pm3$~pc \citet{Kounkel-2018} in the following analysis, as only $100$ of the sources have adopted \textit{Gaia} parallax measurements after the aforementioned quality cut and constraints. Moreover, the uncertainties in \textit{Gaia} parallaxes are too large to be accounted for compared to the size of the ONC. The median uncertainty of the parallax after translated into distance, is $8.7$~pc, whereas the radius of the ONC is only about $3.7$~pc.

We are left with a total number of $235$ sources after applying both the RV and distance constraints: $85$ NIRSPAO sources within $1.52\arcmin$ from the center; $167$ APOGEE sources within $4\arcmin$; $17$ observed with both instruments. 

Note that the proper motion measurements of \textit{HST} + Keck are in the rest frame of the ONC, whereas the \textit{Gaia} DR3 is in the absolute frame. We transform the \textit{Gaia} DR3 proper motion into the same reference frame as in \textit{HST} + Keck by offsetting the former values by the average of their differences in each direction. The offsets are $\left(\widetilde{\Delta\mu_{\alpha^*}},~\widetilde{\Delta\mu_{\delta}}\right) = \left(1.60,~0.08\right) \mathrm{mas}\,\mathrm{yr}^{-1}$. 

Figure~\ref{fig:3d kinematics} visualizes the 3D velocities of the sources\footnote{Note that the proper motions of right ascension are incorrectly labeled in the inverse direction in Figure 8 in \citetalias{Theissen-2022}.}
. The proper motions are characterized by the direction and length of the arrows, while RVs are represented by the color code. Sources moving faster away from us with larger RV values are shown in red, and smaller RVs are shown in blue. Discerning the presence of any kinematic structure solely by visual inspection proves challenging. Therefore, a comprehensive analysis will be conducted in Section~\ref{sec:analysis} to explore this further.

Figure~\ref{fig:kinematics comparison} illustrates the comparison of the kinematics measurements between values adopted in this study and previous observations. The left panel shows the proper motion comparison between \textit{HST} + Keck and \textit{Gaia} DR3. Proper motion in the right ascension and declination are denoted as $\mu_{\alpha^*}$ and $\mu_\delta$, and the measurements from \textit{Gaia} DR3 and \textit{HST} + Keck are denoted as DR3 and HK in the subscript respectively. The source on the bottom left, 2M05351094-0524486 or \textit{Gaia} DR3 3017363547934810112, is the furthest away from consistency because it has a high \texttt{ruwe} of $1.6$, an indication of poor astrometric solution from \textit{Gaia}. In the scope of this work, we only utilize the \textit{Gaia} values if \textit{HST} + Keck RV is not available, as the systematics of the latter are better understood and well accounted for. A similar discussion of comparison between \textit{HST} + Keck and \textit{Gaia} DR2 measurements can be found in \citetalias{Kim-2019}. The right panel shows the RV comparison of the matched sources between NIRSPAO and APOGEE. The binary candidates are marked in different colors.

\begin{figure*}[htb!]
    \centering
    \includegraphics[width=\textwidth]{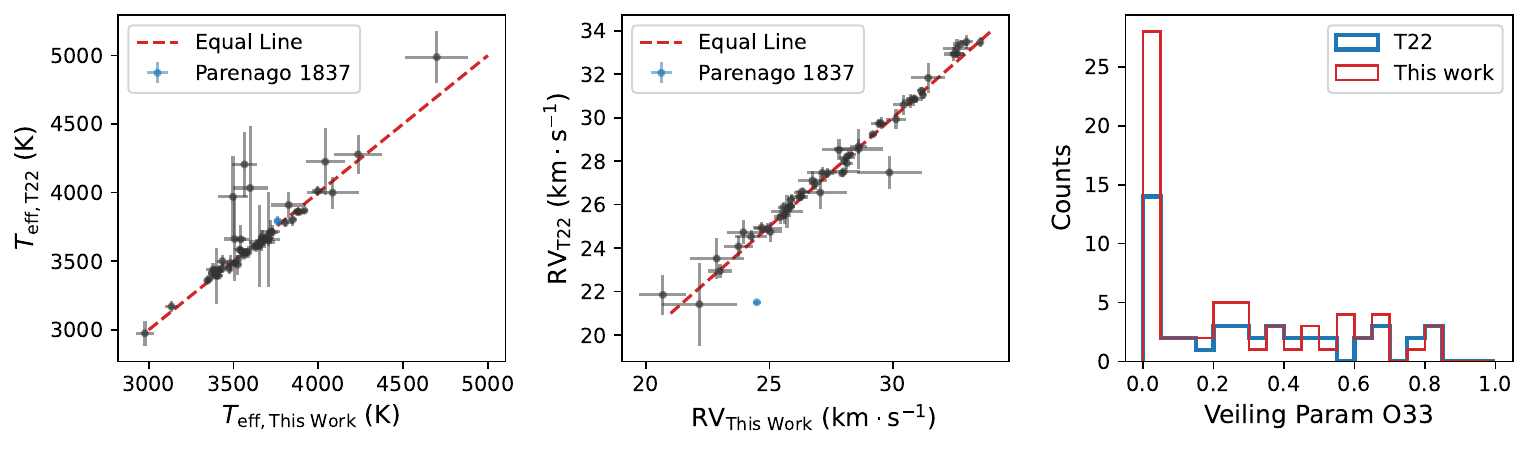}
    \caption{Comparison between the forward-modeled parameters in this work and \citetalias{Theissen-2022}. \textit{Left:} Comparison of effective temperature. The median absolute difference is $29$~K, with a maximum difference of $641$~K. The standard deviation of the differences is $134$~K. \textit{Middle:} Comparison of RV. Note that the most different one shown in blue is the identified binary Parenago 1837. Apart from the binary, the median absolute difference in RV is $0.22~\mathrm{km}\,\mathrm{s}^{-1}$, and the standard deviation of the difference itself is $0.49~\mathrm{km}\,\mathrm{s}^{-1}$. \textit{Right:} Comparison of the distribution of the veiling parameter of order 33.}
    \label{fig:compare T22}
\end{figure*}

The comparison between the forward-modeled parameters of NIRSPAO sources in this work and \citetalias{Theissen-2022} is illustrated in Figure~\ref{fig:compare T22}. Most of the sources have consistent effective temperatures and RVs. The median absolute difference in effective temperature is $29$~K, and the maximum difference is $641$~K, mostly within the error bars of \citetalias{Theissen-2022}. Parenago 1837 accounts for the largest well-constrained RV difference, which is shown in blue. Compared with \citetalias{Theissen-2022}, most of the new sources in this work have low veiling parameters, indicating less dust absorption, which helps disentangle the aforementioned degeneracy between the effective temperatures and veiling parameters, yielding more confidently constrained modeled stellar parameters.

\subsection{Stellar Mass Derivation}
\label{subsec:mass}

\begin{figure}[htb!]
    \centering
    \includegraphics[width=\columnwidth]{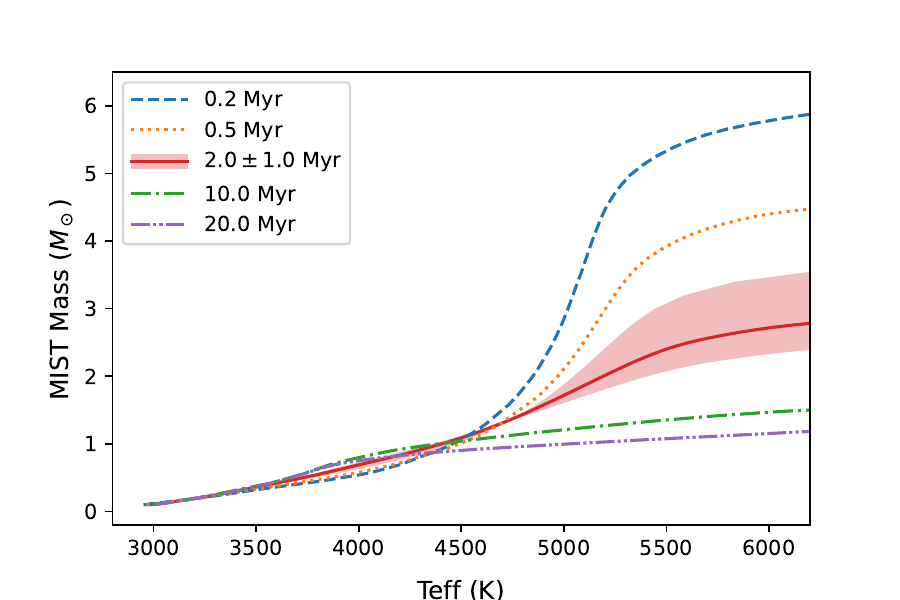}
    \caption{Mass-temperature relation for stars of different ages using the MIST stellar evolutionary model. The red line and shaded area show the model for stars of age $2\pm1$ Myr, which is our assumption for stellar age in this work.}
    \label{fig:mist model}
\end{figure}

With the effective temperature obtained from the forward modeling, stellar mass can be interpolated from evolutionary models. We assume an identical age of $2\pm1$ Myr for all sources in the scope of this work. The reason is twofold. First, the estimated age of the ONC is well-established at around $2$ Myr \citep[][]{Hillenbrand-1997, Reggiani-2011}. Second, the stellar age is only used for the interpolation of stellar mass. Even if we allow a relatively large uncertainty of $50\%$ in the stellar age, it does not affect the mass interpolation for low-mass stars, which are the majority of our sample. Figure~\ref{fig:mist model} illustrates the stellar mass as a function of the effective temperature for stars of different ages under the MIST stellar evolutionary model. Degeneracies do not become significant until the effective temperature exceeds $4500$ K. Indeed, $94\%$ of our sources have a lower effective temperature than $4500$ K. Therefore, it is justified to assume a stellar age of $2\pm1$ Myr for all sources for the purpose of mass interpolation.

Four different stellar evolutionary models are used for the mass interpolation: the MESA Isochrones \& Stellar Tracks (MIST; \citealt{Dotter-2016, Choi-2016}), the BHAC15 model\citep[][]{Baraffe-2015}, non-magnetic isochrones in \citet{Feiden-2016}, and the \citet{Palla-1999} model. The stellar mass is interpolated using the effective temperature under the assumption of a uniform $2\pm1$ Myr age and $0$ metallicity. We calculate the uncertainty in the interpolated mass by taking half of the difference between the highest and lowest mass in the model grid when varying the effective temperature and stellar age within their associated uncertainties, respectively. 

\begin{figure}[htb!]
    \centering
    \includegraphics[width=\columnwidth]{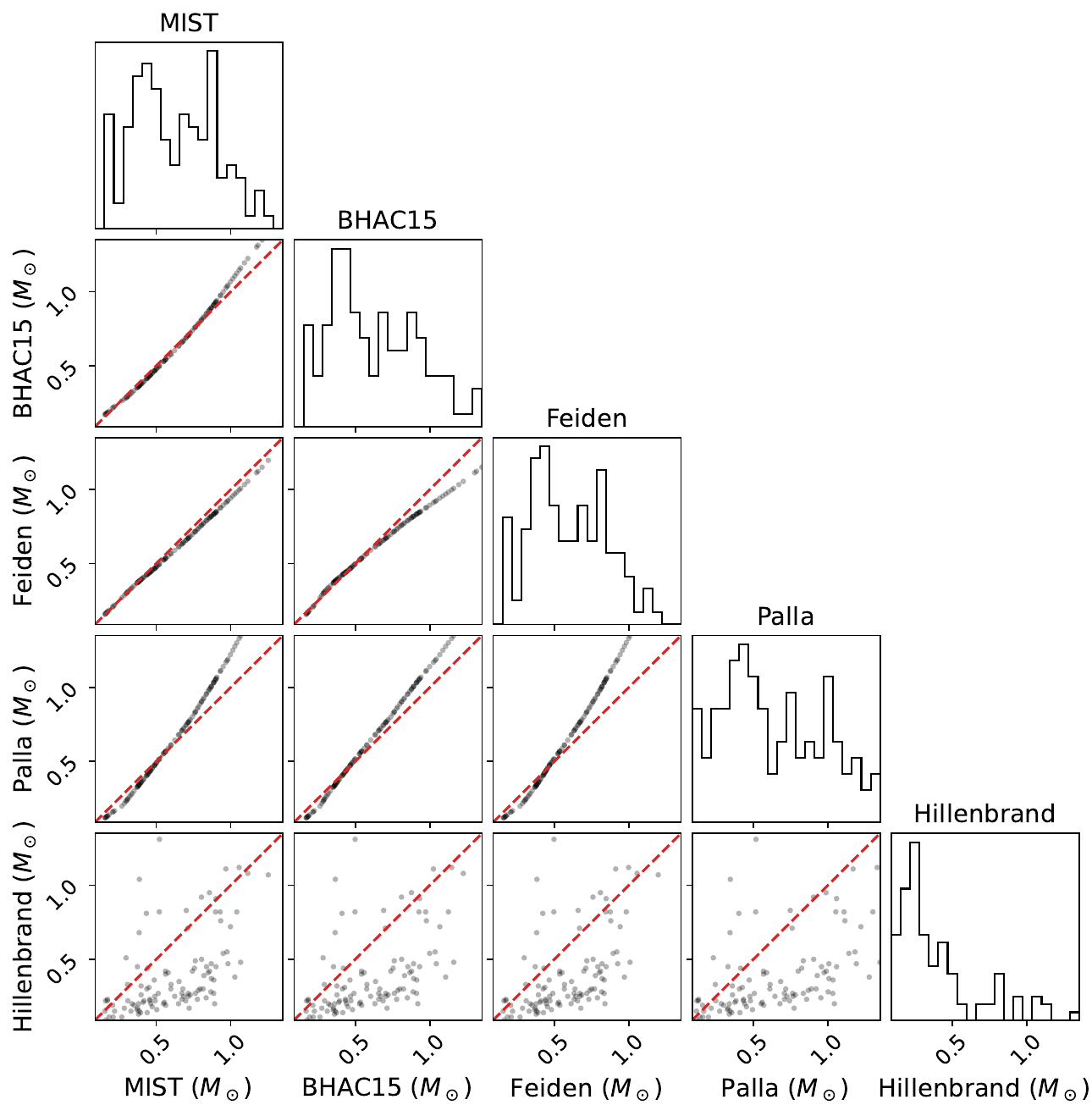}
    \caption{Interpolated stellar mass using four different stellar evolutionary models: MIST, BHAC15, Feiden, and Palla.  Also shown are the values in \citet{Hillenbrand-1997} which uses the evolutionary tracks in \citet{D'Antona-1994} for stars less than $3 M_\odot$.}
    \label{fig:mass comparison}
\end{figure}

Figure~\ref{fig:mass comparison} illustrates the mass comparison using all four models as well as the values from \citet{Hillenbrand-1997}, which uses the evolutionary model by \citet{D'Antona-1994}. Most of the temperature-based interpolated masses agree with one another. However, there is a discrepancy between the masses in \citet{Hillenbrand-1997} and our interpolated values. Effective temperature and luminosity are used to determine the stellar mass in their study. The high nebulosity and reddening in the ONC area make a reliable determination of the luminosity challenging. The use of more recent stellar evolutionary models and age instead of luminosity as the proxy to infer the stellar masses in this work accounts for the difference.

We present the derived parameters from modeling for all sources in our sample in Table~\ref{tab:results}, including RV, $v\sin i$, temperature, and mass.

\section{Mass-Dependent Kinematics Analysis of the Central ONC}
\label{sec:analysis}

\subsection{Virial State and Energy Equipartition}
\label{subsec:virial}


The ONC was previously found to be supervirial (\citealt{Scally-2005, DaRio-2014, Kounkel-2018}; \citetalias{Theissen-2022}). In this study, we revisit this assertion for the central ONC with a larger and denser sample than before in this region.

\citet{DaRio-2014} derived the theoretical velocity dispersion for the ONC based on the stellar and gas density profiles within $3$ pc of the Trapezium center. The stellar density is 
\begin{equation}
    \label{eq:star}
    \rho_\mathrm{star}\simeq70M_\odot\, \mathrm{pc}^{-3} \left(\frac{r}{pc}\right)^{-2.2},
\end{equation}

\noindent where $r$ is the radius from the center of the ONC, and the gas density is

\begin{equation}
    \label{eq:gas}
    \rho_\mathrm{gas}\simeq22M_\odot \, \mathrm{pc}^{-3}.
\end{equation}

\noindent According to the virial theorem, the velocity dispersion $\sigma$ at a certain radius $r$ is related to the enclosed mass within if in virial equilibrium as in Equation~\ref{eq:virial theorem}
\begin{equation}
    \label{eq:virial theorem}
    \frac{GM}{r^2} = 2\sigma^2.
\end{equation}

\noindent Substituting the mass with the integral of the density over the $4$~arcmin (about $0.45$~pc) radius, the dependence of velocity dispersion on the radius can be explicitly derived as in Equation~\ref{eq:vdisp equation}:

\begin{equation}
    \label{eq:vdisp equation}
    \sigma\left(r\right) = \left[\frac{1}{37}\left(\frac{70}{0.8}\left(\frac{r}{pc}\right)^{-0.2} + \frac{22}{3}\left(\frac{r}{pc}\right)^2\right)\right]^{1/2} \mathrm{km}\,\mathrm{s}^{-1}~.
\end{equation}

We measure the velocity dispersion of our sources, assuming there is an intrinsic velocity dispersion along with a measurement uncertainty. That is, the velocity of the $i$-th source in each direction can be parameterized as $v_{(\alpha, \delta, r)_i} \pm \sqrt{\sigma_{{(\alpha, \delta, r)}_i}^2 + \epsilon_{{(\alpha, \delta, r)}_i}^2}$, where $\sigma_{{(\alpha, \delta, r)}_i}$ denotes the intrinsic velocity dispersion and $\epsilon_{{(\alpha, \delta, r)}_i}$ denotes the measurement uncertainty. Note that in this analysis, we excluded sources whose radial velocity deviates more than $3\sigma$ from the mean value to avoid the effects of extreme values, which may be caused by the signal-to-noise ratio of the data or unresolved binaries. The accepted radial velocity range is $27.57\pm13.21~\mathrm{km}\,\mathrm{s}^{-1}$, leaving out $5$ sources with $2$ being Trapezium stars, which are multiple systems. MCMC forward-modeling is adopted to sample the intrinsic velocity dispersion. The same algorithm is used as in \citetalias{Theissen-2022} for the modeling. We directly report our updated values below. The velocity dispersion in each direction and the one-dimensional (1D) velocity dispersion, defined as $\sigma_\mathrm{1D_{3D}}=\sqrt{\left(\sigma_\mathrm{RA}^2 + \sigma_\mathrm{DEC}^2 + \sigma_\mathrm{RV}^2\right)/3}$, are listed in Equation~\ref{eq:vdisp measurements}:
\begin{align} \label{eq:vdisp measurements}
    \sigma_\mathrm{RA} = 1.73 \pm 0.09~\mathrm{km}\,\mathrm{s}^{-1}, \nonumber \\
    \sigma_\mathrm{DEC} = 2.03 \pm 0.11~\mathrm{km}\,\mathrm{s}^{-1}, \nonumber \\
    \sigma_\mathrm{RV} = 2.87 \pm 0.16~\mathrm{km}\,\mathrm{s}^{-1}, \nonumber \\
    \sigma_\mathrm{1D_{3D}} = 2.26 \pm 0.08~\mathrm{km}\,\mathrm{s}^{-1}.
\end{align}
\citet{DaRio-2014} estimate a RMS velocity dispersion in each direction of $1.73$ km/s if the ONC is in virial equilibrium. Our result of $2.26$ km/s is clearly larger than this value (by over $6$--$\sigma$) and indicates that the ONC is supervirial with a virial ratio (kinetic over potential energy) of $q=\left(\sigma_\mathrm{1D_{3D}}/\sigma_\mathrm{equilibrium}\right)^2/2\sim0.85$, consistent with the value in \citet{DaRio-2014}. Therefore, we reconfirm that the ONC center is supervirial.

\begin{figure}[htb!]
    \centering
    \includegraphics[width=\columnwidth]{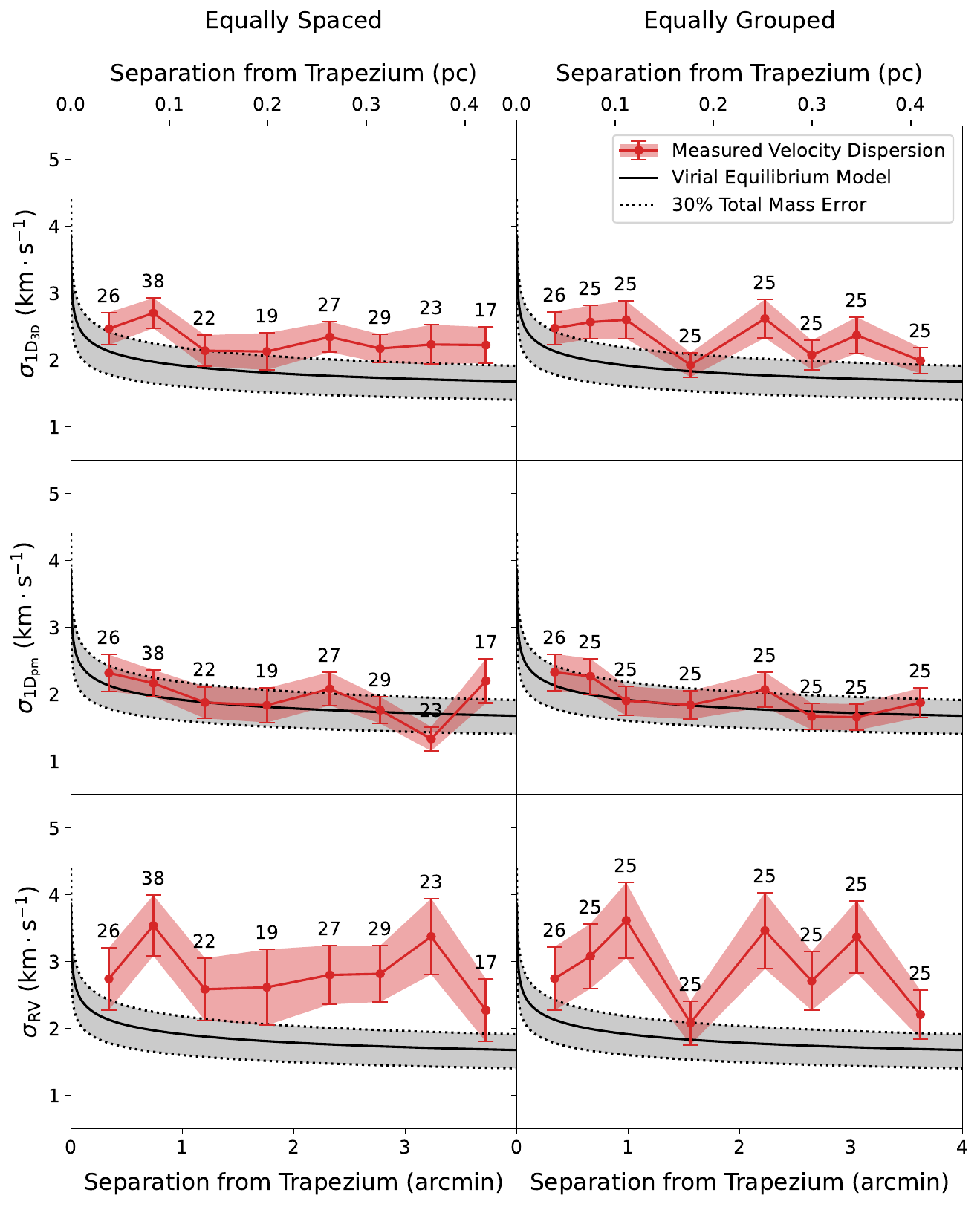}
    \caption{Velocity dispersion as a function of separation from the cluster center. The observed velocity dispersion is shown in red, whereas the requirement of viral equilibrium is illustrated as the black line. The dotted line and the shaded area indicate a $30\%$ total mass error on the virial equilibrium model. The sources are binned equally spaced in the left column, i.e., with identical bin width, and equally grouped in the right column, i.e., with an almost identical number of sources in each bin. The number of sources is labeled on top of each bin. \textit{Top:} 1D velocity dispersion in all directions; \textit{Middle:} 1D velocity dispersion of the proper motions; \textit{Bottom:} Velocity dispersion of the radial velocities.}
    \label{fig:vdisp}
\end{figure}

Figure~\ref{fig:vdisp} shows the velocity dispersion as a function of separation from the center of the cluster. To mitigate against any potential bias in the result due to different ways of binning, the sources are binned equally spaced, i.e., with an identical bin width, and equally grouped, i.e., with an almost identical number of sources in each bin. In Figure~\ref{fig:vdisp}, the left column shows the case of equally spaced bins, while the right column shows equally grouped bins. From top to bottom, each row shows the 1D root-mean-squared velocity dispersion, proper motion component, and radial velocity component, respectively. It can be seen from the middle row in Figure~\ref{fig:vdisp} that the proper motion component is consistent with the virial equilibrium model. However, the radial velocity dispersion (bottom row) is higher than the requirement of virial equilibrium. Consequently, the 1D velocity dispersion of the ONC core sits right above the theoretical prediction of virial equilibrium regardless of the binning method, confirming the result that the ONC is not fully virialized (\citealt{DaRio-2014}; \citetalias{Kim-2019, Theissen-2022}). 


\begin{longrotatetable}
\startlongtable
\begin{deluxetable*}{llcccccccccc}
\tablecaption{NIRSPAO Forward-Modeling Results}
\tabletypesize{\scriptsize}
\label{tab:results}
\tablecolumns{12}
\tablehead{
\colhead{HC2000 ID} & \colhead{K19 ID} & \colhead{APOGEE} & \colhead{Gaia DR3} & \colhead{$\alpha_{J2000}$} & \colhead{$\delta_{J2000}$} & \colhead{$T_\mathrm{eff}$} & \colhead{RV\tablenotemark{a}} & \colhead{$\mu_\alpha\cos\delta$} & \colhead{$\mu_\delta$} & \colhead{\nodata} & \colhead{$M_\mathrm{MIST}$}\\
\colhead{} & \colhead{} & \colhead{} & \colhead{} & \colhead{(deg)} & \colhead{(deg)} & \colhead{(K)} & \colhead{$\left(\mathrm{km}\,\mathrm{s}^{-1}\right)$} & \colhead{$\left(\mathrm{mas}\,\mathrm{yr}^{-1}\right)$} & \colhead{$\left(\mathrm{mas}\,\mathrm{yr}^{-1}\right)$} & \colhead{\nodata} & \colhead{$\left(M_\odot\right)$}
}
\decimals
\startdata
HC2000 322 & --- & --- & 3017364063323188864 & $83.817875$ & $-5.387944$ & $3373.5 \pm 24.8$ & $24.94 \pm 0.57$ & $0.48 \pm 0.13$ & $-0.06 \pm 0.11$ & \nodata & $0.286 \pm 0.021$ \\
HC2000 259 & --- & --- & 3017364059023817600 & $83.821125$ & $-5.392778$ & $3434.2 \pm 33.7$ & $28.02 \pm 0.64$ & --- & --- & \nodata & $0.321 \pm 0.026$ \\
HC2000 213 & --- & --- & 3017363955944598016 & $83.816458$ & $-5.397194$ & $3150.4 \pm 143.9$ & $11.97 \pm 2.70$ & --- & --- & \nodata & $0.167 \pm 0.067$ \\
HC2000 291B & 211 & --- & --- & $83.816000$ & $-5.390444$ & $3171.1 \pm 29.0$ & $29.05 \pm 0.38$ & $1.10 \pm 0.45$ & $-1.63 \pm 0.07$ & \nodata & $0.176 \pm 0.015$ \\
HC2000 252 & 118 & --- & 3017364063325263360 & $83.820750$ & $-5.393639$ & $3162.3 \pm 112.5$ & $24.73 \pm 1.28$ & $2.97 \pm 0.37$ & $2.67 \pm 1.38$ & \nodata & $0.172 \pm 0.054$ \\
HC2000 250 & 197 & --- & 3017363960253005056 & $83.816250$ & $-5.393889$ & $2977.7 \pm 55.1$ & $30.13 \pm 0.42$ & $1.70 \pm 0.40$ & $-3.85 \pm 1.18$ & \nodata & $0.100 \pm 0.012$ \\
HC2000 244 & 180 & --- & 3017363960246048384 & $83.821083$ & $-5.394389$ & $3392.2 \pm 17.7$ & $28.13 \pm 0.32$ & $-0.02 \pm 0.15$ & $0.83 \pm 0.40$ & \nodata & $0.298 \pm 0.016$ \\
HC2000 261 & 206 & --- & 3017363960251927936 & $83.814083$ & $-5.392611$ & $3358.7 \pm 17.7$ & $26.29 \pm 0.20$ & $-0.72 \pm 0.24$ & $0.87 \pm 0.12$ & \nodata & $0.277 \pm 0.017$ \\
HC2000 248 & 200 & --- & --- & $83.815667$ & $-5.394000$ & $3379.4 \pm 40.0$ & $27.06 \pm 1.06$ & $1.81 \pm 0.43$ & $-2.83 \pm 0.20$ & \nodata & $0.290 \pm 0.030$ \\
HC2000 223 & 164 & --- & 3017363960241964032 & $83.814333$ & $-5.395972$ & $3345.9 \pm 20.9$ & $27.35 \pm 0.29$ & $1.10 \pm 0.08$ & $-0.68 \pm 0.23$ & \nodata & $0.270 \pm 0.019$ \\
HC2000 219 & 87 & --- & 3017363960251388288 & $83.813167$ & $-5.396306$ & $3480.2 \pm 39.8$ & $23.75 \pm 0.60$ & $1.87 \pm 0.06$ & $-0.65 \pm 0.02$ & \nodata & $0.348 \pm 0.031$ \\
HC2000 324 & 226 & --- & 3017364028960744704 & $83.811833$ & $-5.387778$ & $3435.8 \pm 29.3$ & $30.70 \pm 0.32$ & $0.70 \pm 0.55$ & $-1.87 \pm 0.04$ & \nodata & $0.322 \pm 0.023$ \\
HC2000 295 & --- & --- & 3017364063325260160 & $83.823208$ & $-5.390250$ & $3506.9 \pm 65.0$ & $20.69 \pm 0.95$ & --- & --- & \nodata & $0.364 \pm 0.047$ \\
HC2000 313 & 198 & --- & 3017364063328048000 & $83.822792$ & $-5.389194$ & $3564.9 \pm 74.1$ & $26.21 \pm 0.37$ & $1.07 \pm 0.86$ & $3.21 \pm 0.20$ & \nodata & $0.398 \pm 0.056$ \\
HC2000 332 & 183 & 2M05353124--0523400 & 3017364059023821696 & $83.824250$ & $-5.387667$ & $3997.3 \pm 36.1$ & $25.45 \pm 0.27$ & $-1.61 \pm 0.17$ & $0.60 \pm 0.38$ & \nodata & $0.685 \pm 0.073$ \\
HC2000 331 & 121 & --- & 3017364063325254528 & $83.826042$ & $-5.387694$ & $3597.6 \pm 103.1$ & $31.43 \pm 0.70$ & $1.26 \pm 0.24$ & $1.03 \pm 0.04$ & \nodata & $0.418 \pm 0.077$ \\
HC2000 337 & --- & 2M05353124--0523400 & 3017364059023825280 & $83.827792$ & $-5.387222$ & $4652.1 \pm 284.4$ & $31.91 \pm 0.28$ & $-0.22 \pm 0.04$ & $0.07 \pm 0.04$ & \nodata & $1.252 \pm 0.415$ \\
HC2000 375 & 560 & 2M05353124--0523400 & 3017364063330453504 & $83.820750$ & $-5.383611$ & $4040.9 \pm 115.3$ & $32.60 \pm 0.56$ & $0.78 \pm 0.13$ & $1.15 \pm 0.04$ & \nodata & $0.716 \pm 0.131$ \\
HC2000 388 & 532 & --- & 3017364063333454336 & $83.823167$ & $-5.382444$ & $4234.5 \pm 141.4$ & $30.44 \pm 0.44$ & $0.65 \pm 0.23$ & $0.35 \pm 0.04$ & \nodata & $0.857 \pm 0.163$ \\
HC2000 425 & --- & --- & 3017364127743299328 & $83.824792$ & $-5.379306$ & $3546.2 \pm 20.3$ & $24.69 \pm 0.33$ & $2.37 \pm 0.03$ & $2.67 \pm 0.03$ & \nodata & $0.387 \pm 0.022$ \\
$\vdots$ & $\vdots$ & $\vdots$ & $\vdots$ & $\vdots$ & $\vdots$ & $\vdots$ & $\vdots$ & $\vdots$ & $\vdots$ & $\vdots$ & $\vdots$\\
\enddata
\tablenotetext{a}{Heliocentric corrected RV}
\tablecomments{Only a portion of the table is shown here. A complete version of this table is available in the online version of this paper.}
\end{deluxetable*}
\end{longrotatetable}
\clearpage

\begin{figure*}[htb!]
    \centering
    \includegraphics[width=\textwidth]{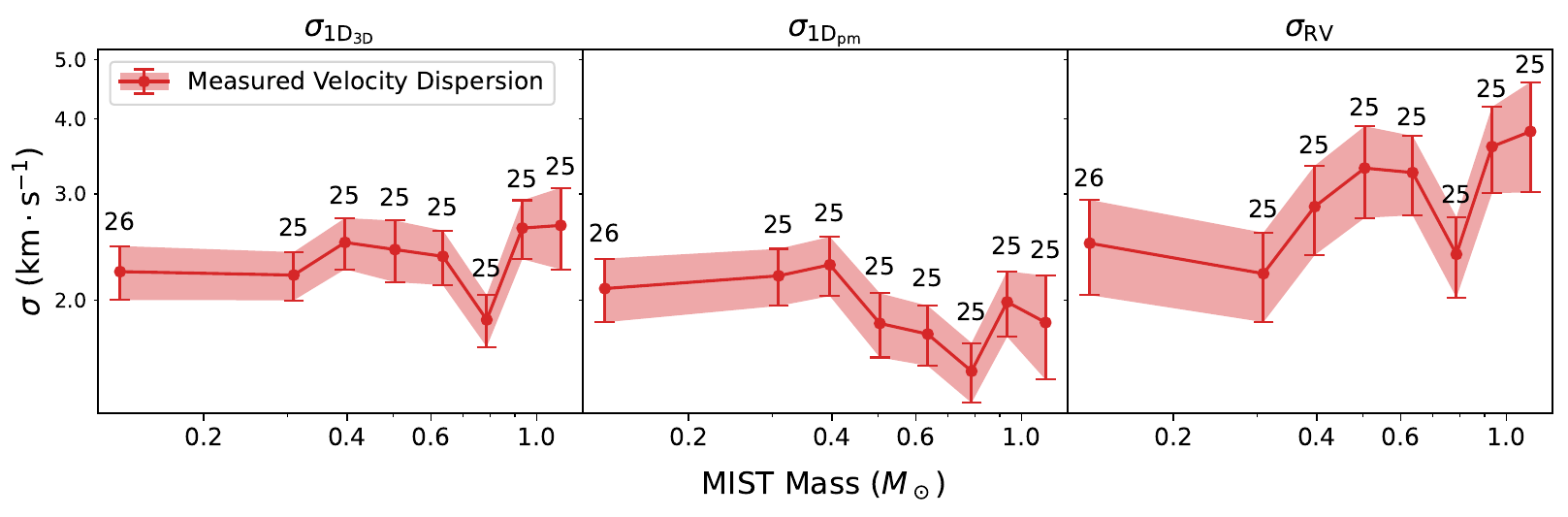}
    \caption{Velocity dispersion as a function of stellar mass interpolated from the MIST model. From left to right, the three subfigures show the 1D velocity dispersion of all three directions $\sigma_\mathrm{1D_{3D}}$, the 1D velocity dispersion of the proper motions $\sigma_\mathrm{1D_{pm}}$, and the radial velocity dispersion $\sigma_\mathrm{1D_{RV}}$ respectively. The sources are grouped with equal sizes, and the number of sources in each bin is labeled on top of the corresponding bin.}
    \label{fig:vdisp vs mass}
\end{figure*}

Another dynamical state we can infer from the velocity dispersion is whether energy equipartition has occurred in the cluster. The velocity dispersion should be inversely proportional to the square root of the stellar mass in a cluster where energy equipartition has already occurred via gravitational interactions. Previously, \citet{Hillenbrand-1998} did not see evidence of equipartition in the ONC. Here we re-evaluate this conclusion with our newest data. Figure~\ref{fig:vdisp vs mass} shows the equally grouped 1D velocity dispersion of all three directions $\sigma_\mathrm{1D_{3D}}$, 1D velocity dispersion in the proper motion directions $\sigma_\mathrm{1D_{pm}} = \sqrt{\left(\sigma_\mathrm{RA}^2 + \sigma_\mathrm{DEC}^2\right)/2}$, and radial direction as a function of the stellar mass. The $-1/2$ slope is clearly not present. This is conceivable as the relaxation time of the cluster is estimated to be $6.5$~Myr \citep[][]{Hillenbrand-1998}, much larger than the age of the cluster of about $2$~Myr. Therefore, energy equipartition has not yet taken place in the central ONC, in agreement with \citet{Hillenbrand-1998}.

\subsection{Velocity-Mass Relation}
\label{subsec:v-m}

\begin{figure*}[htb!]
    \centering
    \gridline{\fig{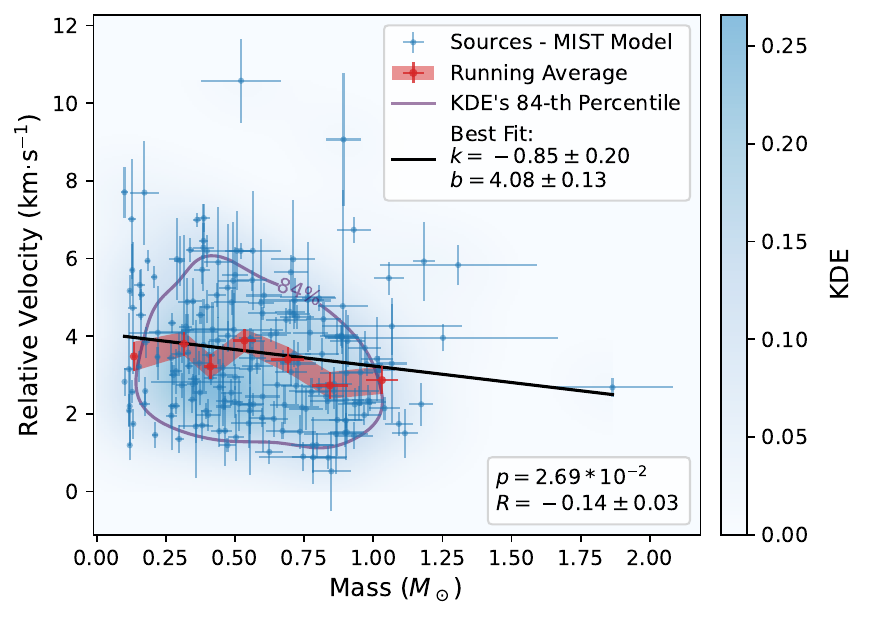}{0.5\textwidth}{(a)}
              \fig{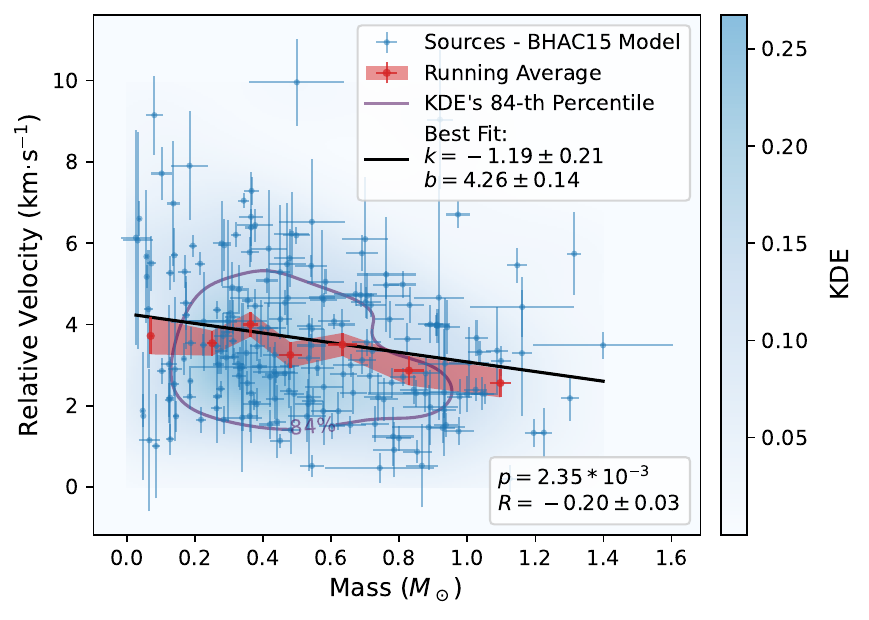}{0.5\textwidth}{(b)}}
    \gridline{\fig{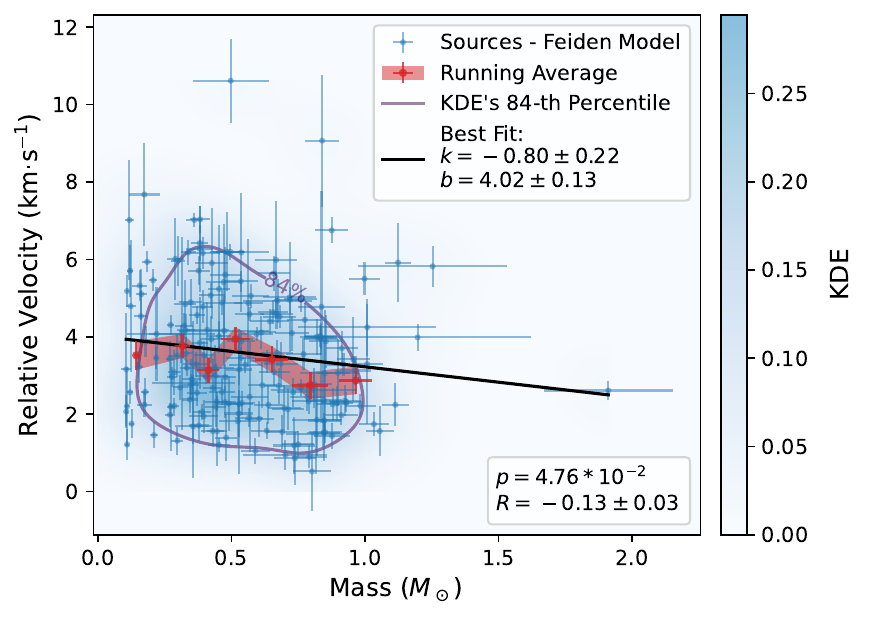}{0.5\textwidth}{(c)}
              \fig{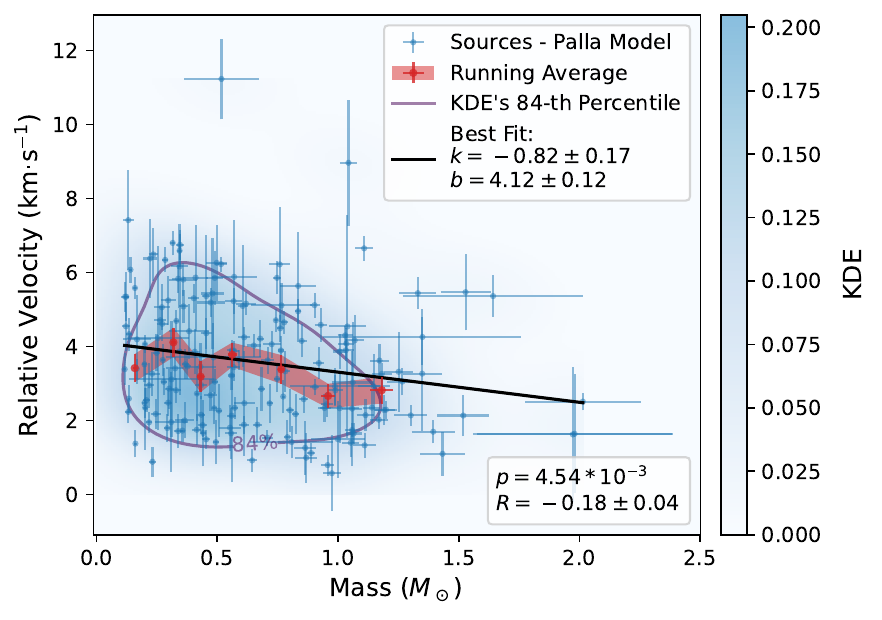}{0.5\textwidth}{(d)}}
    \caption{The computed velocity relative to the center of mass of the neighbors of each source within a $0.1$-pc (or $53\arcsec$) radius versus stellar mass. Four models are used to interpolate the stellar mass: (a) MIST model; (b) BHAC15 model; (c) Feiden model; (d) Palla model. In each subfigure, the data are represented as blue points and error bars. The value of the kernel density estimator (KDE) is colored in blue in the background to show the distribution of the data. The purple line marks the $84$-th percentile of the KDE value. The equally-grouped running average, along with its uncertainties, is marked and filled in red. The black line shows the best linear fit to the data, with the slope $k$ and intercept $b$ labeled in the legend. $p$-value and Pearson's correlation coefficient $R$ are labeled in the bottom right of each figure. All four models display both a negative slope of the linear fit and a negative correlation coefficient.}
    \label{fig:vrel-mass}
\end{figure*}


With kinematic information and mass estimates for sources in the central ONC, we can look for correlations between masses and velocities, which is indicative of whether stars form via filament fragmentation. Figure~\ref{fig:vrel-mass} shows the velocity of each source relative to its neighbors (including itself) within a $0.1$~pc (or $53\arcsec$) radius versus their masses derived from the four different evolutionary models described in Section~\ref{subsec:mass}. The radius within which sources are considered as neighbors can be varied and will be further discussed in Section~\ref{subsec:teff offset}. The median number of neighbors for each source is $11$. The data are shown in blue. We employ the Gaussian kernel density estimation (KDE) to visualize the distribution of the data, and the value of the estimator is colored in blue in the background. The $84$-th percentile of the estimator value is marked as the purple curve. As can be seen, the envelope displays a negative trend on its upper edge. To look at this trend more globally, we show the equally grouped running average in red.  Since the trend of velocity versus mass appears roughly linear, we fit a linear relationship $v_\mathrm{rel}=k*m+b$ to the data using the \texttt{scipy.stats.linregress} function \citep[][]{scipy}, where $v_\mathrm{rel}$ is the relative velocity, $m$ is the stellar mass, $k$ and $b$ are the slope and intercept, respectively.  To determine the best-fit slope and intercept and their associated uncertainties, we resample the relative velocity and the mass of each source from a normal distribution centered at the observed values, with standard deviations being the uncertainty of measurement. Only positive values are kept for the linear fitting. We resampled $10^5$ times and each time conducted a linear regression on the data. The median of the recorded slope and intercept is chosen as the best fit values, while half the differences between the $16$ and $84$-th percentiles are considered as their associated uncertainties. The values and uncertainties of the slope $k$ and the intercept $b$ are labeled in the legend in each figure. We utilize the measured values to calculate the $p$-value for the linear fit, with the null hypothesis being that the slope is zero. As all of the $p$-values are less than $0.05$, we can safely reject the null hypothesis and conclude that the negative correlation is statistically significant. Apart from the linear fit, we also calculate the Pearson correlation coefficient $R$ as a reflection of the degree of their correlation.  The best-fit value and uncertainty of $R$ are determined by resampling in the same way as the linear fit.  As can be seen from the figure, all four models display a negative correlation. 

\begin{figure}
    \centering
    \includegraphics[width=\columnwidth]{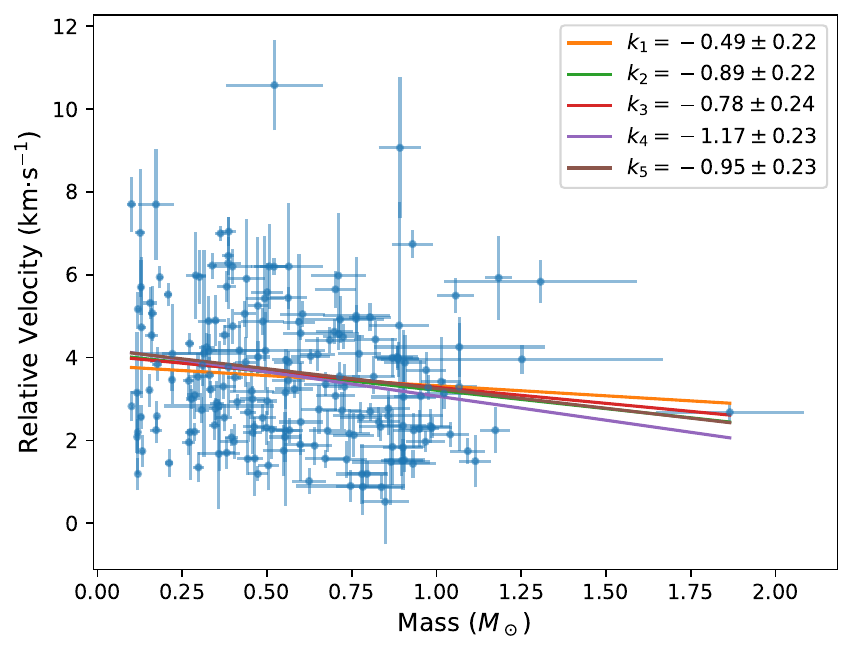}
    \caption{$5$-fold cross-validation of the negative correlation between relative velocity and stellar mass. The slope and its associated uncertainty for each linear regression $k_1$--$k_5$ is labeled in the legend. This test adopted the stellar masses derived from the MIST model.}
    \label{fig:kfold}
\end{figure}

We utilized $5$-fold cross-validation to verify the statistical significance of the negative correlation between relative velocity and stellar mass and ensure that it is not driven by only a few outlier datapoints. The data is randomly partitioned into $5$ equally sized groups, commonly referred to as `folds'. Linear regression is then performed on $4$ groups of the data, leaving out a different group each time. The slope of the resulting $5$ linear regressions in the case of the MIST model is shown in Figure~\ref{fig:kfold}. The uncertainty-weighted average of the slope across the $5$ folds is $-0.85$, and the standard deviation is $0.22$, consistent with the result shown in Figure~\ref{fig:vrel-mass} (a). Therefore, by randomly selecting $5$ different sets of $80\%$ of the data but still arriving at the same conclusion, we further validated our finding that the relative velocity is negatively correlated with stellar mass.



\subsection{Effective Temperature Offset Between NIRSPAO and APOGEE}
\label{subsec:teff offset}
\begin{figure}
    \centering
    \includegraphics[width=\columnwidth]{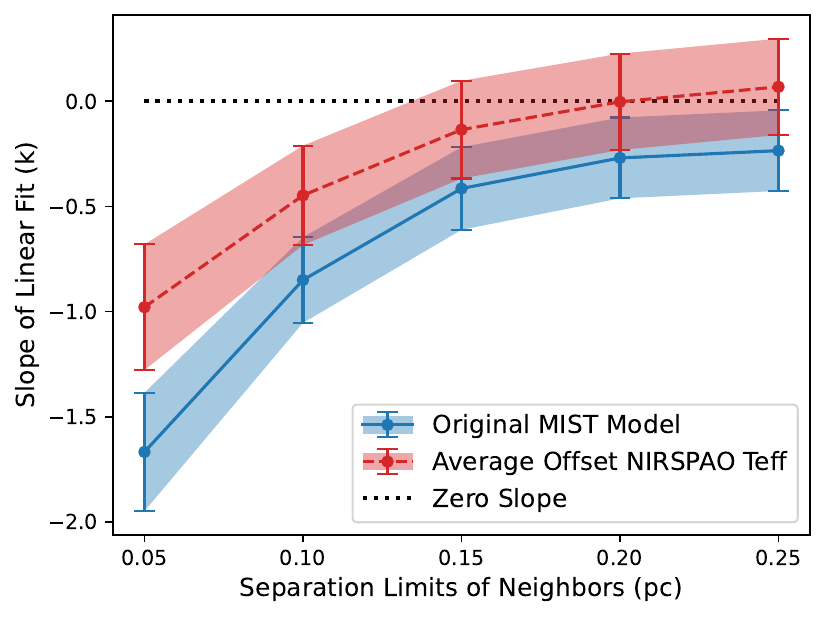}
    \caption{Slope of linear fit as a function of separation limit of neighbors. We selected $0.1$~pc when analyzing the velocity-mass relation in Section~\ref{subsec:v-m}. Here, the separation limit within which sources are counted as neighbors is varied to explore how the linear slope changes. The blue line shows the stellar mass under the MIST model with the derived $T_\mathrm{eff}$, while the red line shows the MIST stellar mass after offsetting the NIRSPAO $T_\mathrm{eff}$ by the average difference of $526.25$~K between NIRSPAO and APOGEE. The dashed line shows the zero-slope, or no correlation. The negative correlation between relative velocity and mass is more evident locally within a smaller radius when calculating the relative velocity.}
    \label{fig:slope_vs_mass}
\end{figure}

Despite using the same spectral modeling algorithm, there seems to be a systematic discrepancy in the effective temperatures between the $17$ cross-matched NIRSPAO and APOGEE sources in our sample. Figure~\ref{fig:slope_vs_mass} shows the comparison. Effective temperatures derived from APOGEE spectra are higher than NIRSPAO results, with a weighted average offset of $586$~K and a maximum difference of $892$~K. Several reasons may cause this difference. First, the spectra of APOGEE and NIRSPEC are in $H$ and $K$ bands, respectively. The spectral features that determine the modeled stellar parameters are therefore different between the two sets of observations. For example, the CO lines are more sensitive to low-temperature sources, as the intended objective of the NIRSPAO observation is to observe low-mass stars. Additionally, considering the highly embedded and crowded nature of the region, extinction and reddening are not identical in $H$ and $K$ bands. A future study will simulate the effect of reddening on temperature estimates. Additionally, to enhance the quality of modeling, especially RV, there is an ongoing effort to model the fringing in the spectrum, which is the primary contributor to the residuals.  For the purpose of this study, NIRSPAO estimated effective temperatures are prioritized over APOGEE results.

To evaluate the impact of choosing the NIRSPAO temperatures for mass estimates, we offset the NIRSPAO temperature by the weighted-averaged difference of $586$~K to simulate the effect on the velocity-mass relation discussed in Section~\ref{subsec:v-m}. Figure~\ref{fig:slope_vs_mass} shows the slope of the linear fit before and after the offset as a function of the radius within which sources are considered as neighbors when calculating relative velocities. It can be seen that the negative correlation with mass is weaker but still exists after inflating the NIRSPAO temperature to match the APOGEE values. Either before or after the offset, the negative trend between relative velocity and mass is more evident locally, i.e., a smaller threshold of radius for sources to be considered as neighbors. Discretion is advised for the near-zero slope at larger neighboring radius, as the entire area being analyzed is about $0.45$~pc in radius. More sources on the periphery would have incomplete neighbors, which could affect the accuracy of the correlation and the underlying significance.

\subsection{Preferred Proper Motion Direction}
\label{subsec:pm}

\begin{figure}
    \centering
    \includegraphics[width=\columnwidth]{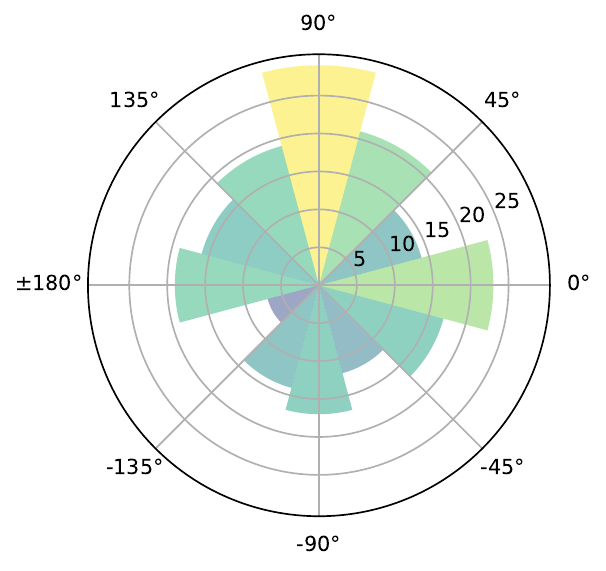}
    \caption{Distribution of the angle between the displacement vector from the center of the ONC and the proper motion vector on the plane of the sky. Positive values stand for clockwise rotation about the center of the ONC, and vice versa. $0^\circ$ stands for expansion and $\pm180^\circ$ means contraction towards the center. More sources are in the $0^\circ$ than $\pm180^\circ$ bin agrees with the finding that the ONC is experiencing a slight expansion \citep[][]{Kounkel-2022}. The peak at $90^\circ$ here illustrates that the sources have a preference for clockwise rotation on the plane of the sky around the ONC center.}
    \label{fig:pm direction}
\end{figure}

Previous studies identified signs of expansion within the ONC \citep[e.g.,][]{Kounkel-2022}, and a rotational preference in the proper motions \citepalias{Theissen-2022}. With a combination of HST and Gaia measurements, we are able to re-evaluate the above findings in greater detail. A polar histogram of the angle between the stellar proper motion vector and the separation vector from the ONC center is shown in Figure~\ref{fig:pm direction}. Specifically, a positive angle represents clockwise rotation about the ONC center, whereas a negative angle represents counter-clockwise rotation. An angle of zero indicates the source is moving radially outward with respect to the ONC center on the plane of the sky, while $-180^\circ$ means the source is moving towards the center. More sources are in the $0^\circ$ than $\pm180^\circ$ bin agrees with the finding that the ONC is experiencing a slight expansion \citep[][]{Kounkel-2022}. The peak around $90^\circ$ in Figure~\ref{fig:pm direction} illustrates that the center of the cluster is undergoing a clockwise rotation. This is consistent with the finding in \citet{Strand-1958}. \citetalias{Theissen-2022} identifies a rotational preference in both $+90^\circ$ and $-90^\circ$, or clockwise and counter-clockwise directions simultaneously. The increase in sample size shows that the ONC core is actually experiencing a clockwise rotation, though a larger sample size would help confirm this trend.

\section{Binary Candidates and Binary Simulation}
\label{sec:binary}
Among the $23$ sources that have multiple epochs of RV measurements, $4$ of them exhibited strong variability in their radial velocities. We therefore report the discovery of two candidate binary systems, Paranego 1837, Brun 590, V* V1337 Ori, and V* V1279 Ori.

\subsection{Parenago 1837}
\label{subsec:parenago}
\begin{figure}
    \centering
    \includegraphics[width=\columnwidth]{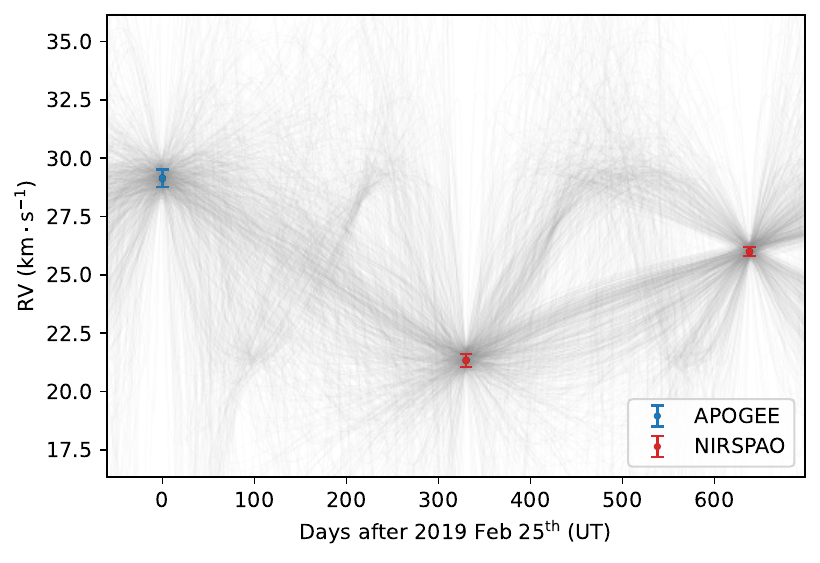}
    \caption{The RVs after barycentric correction of the identified binary candidate Parenago 1837 measured by APOGEE and NIRSPAO in three epochs. $876$ potential orbital fits with a period between $\Delta t/3$ ($213$~days) and $2\Delta t$ ($1276$~days) are shown in gray. Three different modes of orbits can be clearly seen from the figure, corresponding to periods within ranges of $\Delta t/4\sim\Delta t/2$, $\Delta t/2\sim\Delta t$, and $\Delta t\sim2\Delta t$, respectively.}
    \label{fig:orbital fits}
\end{figure}

\begin{figure}
    \centering
    \includegraphics[width=\columnwidth]{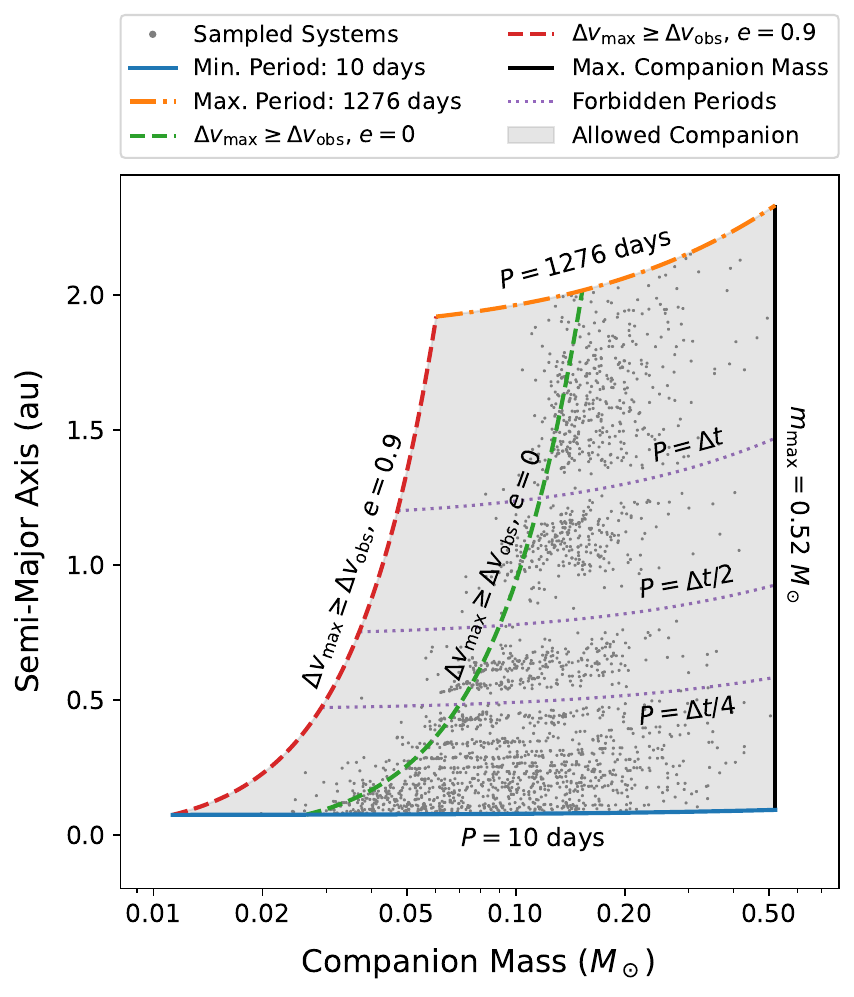}
    \caption{Allowed companion in the semi-major axis-companion mass parameter space and $2126$ sampled systems from \textit{The Joker}. Each sampled system is shown as a small gray point. The period range between $10$~days and twice the observation time span $2\Delta t$ or $1276$~days is indicated by the blue solid line at the bottom and the amber dash-dotted line at the top, respectively. The left boundary is set by the required variation in orbital velocity, which is illustrated in green dashed line for circular orbits, and in red for orbits with eccentricity up to $0.9$ (see Equation~\ref{eq:deltav}). Assuming a companion smaller than the observed primary mass of $0.52~M_\odot$ sets the limit shown by the dotted black line to the right. The shaded area is the allowed parameter space within which the companion can reside with the assumptions above. The forbidden periods, which are an integer fraction of the observation time span $\Delta t$, are labeled as the purple dotted line.}
    \label{fig:allowed param}
\end{figure}

Parenago 1837, or HC2000 546, exhibits variation in its RVs measured in three different epochs, first by APOGEE, followed by two observations by NIRSPAO. According to \textit{Gaia} DR3, it has a G magnitude of $13.54$, with BP (blue pass) magnitude of $13.23$ and RP (red pass) magnitude of $11.95$. The derived stellar mass of the primary is $0.52\pm0.04~M_\odot$ according to the MIST model, assuming the primary light dominates the observed spectra. The RVs after barycentric correction are $29.14\pm0.38\mathrm{km}\,\mathrm{s^{-1}}$, $21.34\pm0.28~\mathrm{km}\,\mathrm{s^{-1}}$, and $25.99\pm0.19~\mathrm{km}\,\mathrm{s^{-1}}$ measured on on 2019 February 25$^\mathrm{th}$ (UT), 2020 Janurary 21$^\mathrm{st}$ (UT), and 2021 October 20$^\mathrm{th}$ (UT), respectively. The total time span between the first and last observation $\Delta t=638$ days. Figure~\ref{fig:orbital fits} shows each measured RV at the time of observation. The blue and red error bars indicate the results from APOGEE and NIRSPAO, respectively.

To explore the possible properties of the companion, we sample the possible orbital parameters of the system using a Monte Carlo rejection sampler, \textit{The Joker} \citep{thejoker}. \textit{The Joker} requires specifying the priors, including the period limits, the RV semi-amplitude $K$, and the standard deviations of the velocity trend priors. We limit the periods to between $10$ days and twice the observation time span $\Delta t$, or $1276$ days. Orbital solutions with arbitrarily long periods and large RV variations can be obtained with only $3$ epochs of observations. Therefore, an upper limit on the period is a reasonable assumption considering the low mass of the object. The semi-amplitude prior is set to $5~\mathrm{km}\,\mathrm{s}^{-1}$, slightly larger than the variation of the RVs. The standard deviations of the velocity priors are set to be a relatively large value of $100~\mathrm{km}\,\mathrm{s}^{-1}$ to allow the sampler to fully explore the orbital parameter space. Additionally, we replace the default prior distribution of eccentricity with a uniform distribution between $0$ and $0.9$ to evenly explore the parameter space. We generate $10^5$ prior samples and $2126$ orbital solutions that match the observed RVs remain after the rejection sampling. The semi-amplitude of the primary's RV variation $K$ is related to stellar masses as
\begin{equation}
\label{eq:K_original}
    K=\frac{m}{M+m}\frac{2\pi a\sin I}{T\sqrt{1-e^2}}~,
\end{equation}
where $M$ is the primary mass, $m$ is companion mass, $T$ is the orbital period, $a$ is the semi-major axis, $I$ is the inclination, and $e$ is the eccentricity of the orbit \citep{Murray-2010}. The semi-major axis $a$ can be expressed as the stellar masses and the orbital period $T$ according to Kepler's third law
\begin{equation}
\label{eq:a}
    a=\left(\frac{\mu}{4\pi^2}T^2\right)^{1/3}~,
\end{equation}
where $\mu=G\left(M+m\right)$ is the standard gravitational parameter. Substituting Equation~\ref{eq:a} into Equation~\ref{eq:K_original}, $K$ can be expressed as a function of stellar masses, eccentricity, and orbital period
\begin{equation}
\label{eq:K}
    K=\frac{m}{M+m}\frac{2\pi\sin I}{\sqrt{1-e^2}}\left(\frac{\mu}{4\pi^2T}\right)^{1/3}~.
\end{equation}
Therefore, we can solve for the minimum companion mass when $I=\pi/2$ given the $K$, $T$, and $e$ for each sampled system from Equation~\ref{eq:K} with the derived primary mass of $0.50~M_\odot$. Figure~\ref{fig:allowed param} shows the distribution of sampled systems and the theoretically allowed region in the semi-major axis-companion mass parameter space. The bottom and top limits are set by the assumed period range between $10$ days and $1276$ days. The left boundary is set by the constraint that the maximum variation in orbital velocity difference when the orbit is observed from an edge-on direction must exceed the observed amplitude of RV change, i.e., 
\begin{equation}
\label{eq:deltav}
\begin{split}
\Delta v_\mathrm{max}   & = v_p + v_a \\
                        & = \frac{m}{M+m}\sqrt{\frac{\mu}{a}}\left(\sqrt{\frac{1+e}{1-e}} + \sqrt{\frac{1-e}{1+e}}\right) \\
                        & = 2\frac{m}{M+m}\sqrt{\frac{\mu}{a\left(1-e^2\right)}} \\
                        & \geq\Delta v_\mathrm{obs}~,
\end{split}
\end{equation}
where $v_\mathrm{max}$ is the maximum orbital velocity difference, $\Delta v_\mathrm{obs}$ is the observed RV difference, and $v_p$, $v_a$ denotes the orbital velocity at perihelion and aphelion, respectively \citep{Murray-1999}. Both the cases when $e=0$ and $e=0.9$ are plotted in Figure~\ref{fig:allowed param} in green and red dashed lines. As can be seen, most sampled systems are within the green line, with a small fraction residing between the green and the red line. Assuming a companion smaller than the observed primary mass of $0.50~M_\odot$ sets the limit shown by the dotted black line to the right. The gaps visible in the sampled systems and marked by purple dotted lines are the forbidden periods of integer number fractions of the observation time span $\Delta t$, as the first and last measured RVs are not identical.

$876$ potential orbital fits with a period greater than $\Delta t/3$ or $213$ days but less than $2\Delta t$ or $1276$ days are plotted in gray in Figure~\ref{fig:orbital fits}. Three different modes of orbits can be clearly seen from the figure, corresponding to periods within ranges of $\Delta t/4\sim\Delta t/2$, $\Delta t/2\sim\Delta t$, and $\Delta t\sim2\Delta t$, respectively.

Despite the limited epochs of observation, we are able to infer the approximate mass and separation of the companion thanks to the derived primary mass under reasonable assumptions on the orbital period between $10$ days and twice the observation time span of $1276$ days. Parenago 1837 is a candidate binary system consisting of a primary of $0.52\pm0.04~M_\odot$ and most likely a companion of $\sim0.03$--$0.3~M_\odot$, with a separation of less than $2$~au. Further observation is needed to more robustly constrain its orbit.

\subsection{V* V1337 Ori}
\label{subsec:V1337}
V* V1337 Ori, or HC2000 214, is another binary candidate. The original APOGEE results have $6$ visits on the object. The difference in RV is as large as $17.87~\mathrm{km}\,\mathrm{s}^{-1}$ throughout all the visits, which makes it very likely to be a binary system. However, currently we only have one reanalyzed APOGEE result. Single-epoch reanalysis is required to further constrain its orbit. For consistency, we utilize the reanalyzed APOGEE result of $45.52\pm1.25~\mathrm{km}\,\mathrm{s}^{-1}$, followed by NIRSPAO measurement on 2020 January 20$^\mathrm{th}$ (UT) of $36.93\pm1.05~\mathrm{km}\,\mathrm{s}^{-1}$. The two measurements differ by more than $7$-sigma. The inferred primary mass is $0.52\pm0.14~M_\odot$. Due to the limited measurements, it would be challenging to do a similar analysis of Parenago 1837. Therefore, reanalysis on individual previous APOGEE visits or future observations is needed to confirm and constrain this binary system.

\subsection{V* V1279 Ori}
V* V1279 Ori, or HC2000 170, is a source of $0.40\pm0.03~M_\odot$ according to the MIST model. It has a RV measurement of $23.7\pm1.0~\mathrm{km}\,\mathrm{s}^{-1}$ in the RV survey by \citet{Tobin-2009}. Our Keck NIRSPAO measures a RV of $32.70\pm1.62~\mathrm{km}\,\mathrm{s}^{-1}$ on 2022 Jan 18$^\mathrm{th}$ (UT), more than $5$--$\sigma$ different from each other.

\subsection{Brun 590}
Brun 590, or HC2000 172, is another candidate of $0.60\pm0.06~M_\odot$ according to the MIST model with $2$ RV measurements. 
NIRSPAO measures $23.36\pm0.29~\mathrm{km}\,\mathrm{s}^{-1}$ on 2022 Jan 20$^\mathrm{th}$ (UT). Different interpretations of APOGEE RV is present in the literature. The reanalyzed APOGEE RV from \citetalias{Theissen-2022} is $29.26\pm1.23~\mathrm{km}\,\mathrm{s}^{-1}$, while \citet{Kounkel-2019} gives $19.320\pm1.182~\mathrm{km}\,\mathrm{s}^{-1}$ after removing the systematic effect of temperature and epoch-dependent offsets \citep[][]{Cottaar-2014}. Additional validation and observation is required to confirm whether it is a binary system and constrain the orbit.

\subsection{Binary Simulation}
\label{subsec:binary simulation}

\subsubsection{Velocity Dispersion}
\begin{figure}[htb!]
    \centering
    \includegraphics[width=\columnwidth]{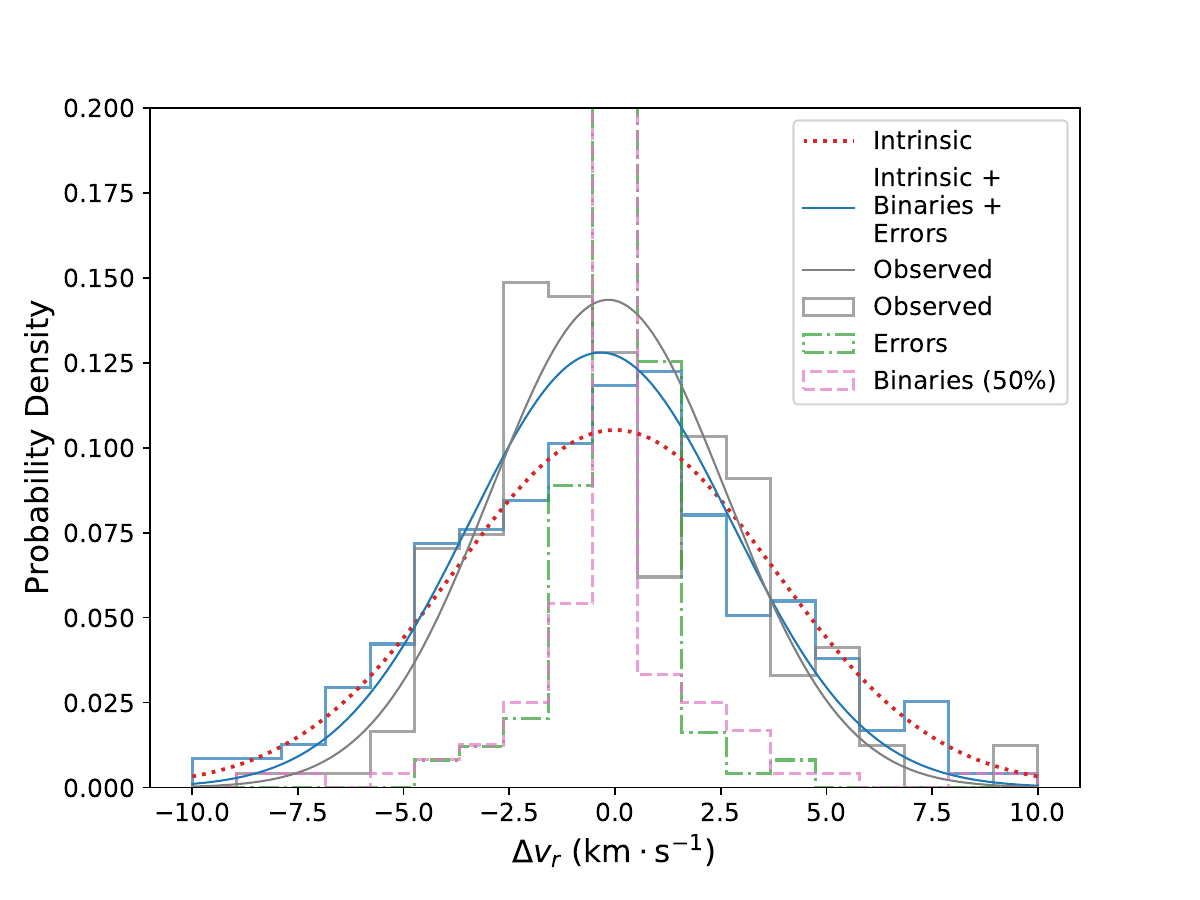}
    \caption{Radial velocity dispersion in a simulation with binary fraction set to $50\%$. The red dotted line shows the intrinsic velocity dispersion. The gray histogram and curve show the observed distribution and fitted normal distribution. The synthetic velocity dispersion with contributions from the intrinsic velocity dispersion, the measurement errors, and the binary offset is shown in blue. The dash-dotted green histogram shows the measurement errors, and the pink dashed histogram shows the contribution from the binaries.}
    \label{fig:binary histogram}
\end{figure}

\begin{figure}[htb!]
    \centering
    \includegraphics[width=\columnwidth]{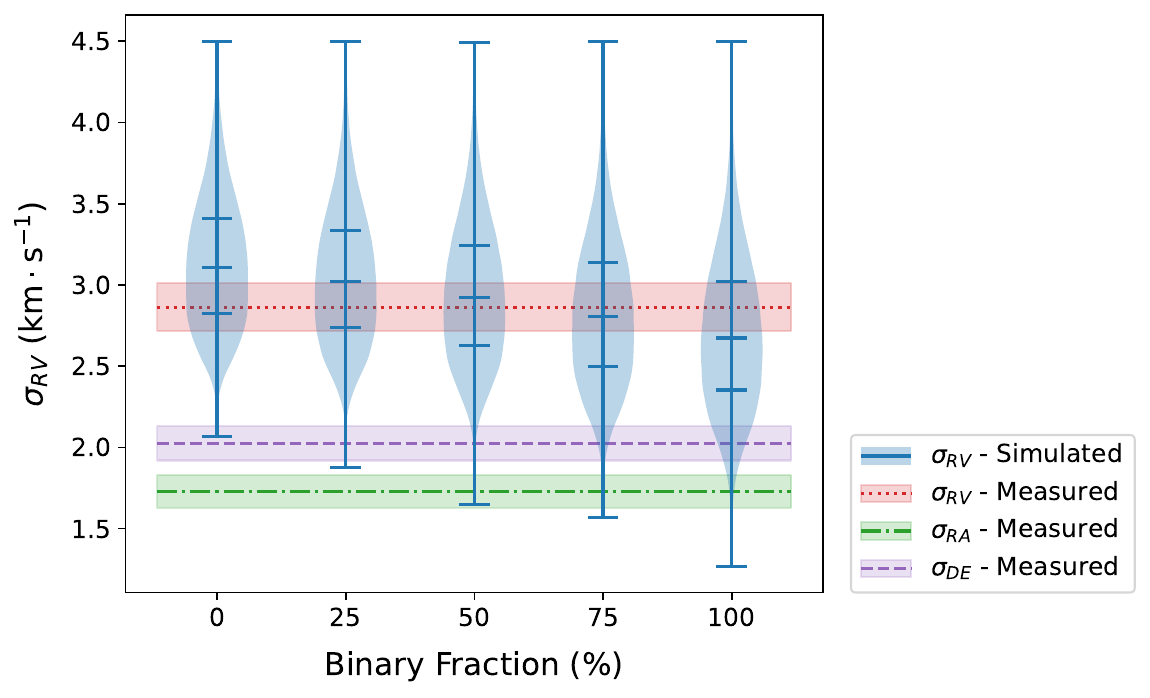}
    \caption{Binary simulation result of the intrinsic radial velocity dispersion. The blue violin plot shows the distribution of the $10^5$ simulated synthesized radial velocity dispersion under different imposed binary fractions, with the $0$-th, $16$-th, $50$-th, $84$-th, and $100$-th percentiles marked by the horizontal bars. The measured velocity dispersion in radial, right ascension, and declination directions is shown in red dotted line, green dash-dotted line, and purple dashed lines, respectively.}
    \label{fig:binary violin}
\end{figure}

The Keck can spatially resolve binaries with separations larger than $25$~mas, or about $10$~au at the distance of the ONC \citep{Lacour-2011}. Closer binary systems can not be resolved. Currently, we do not have a measurement of the close binary fraction in the ONC, as most of the published work has focused on visual binaries.  A large fraction of optically unresolved binaries will have a profound effect on the IVD. Therefore, we use the \texttt{velbin} package \citep[][]{Cottaar-2014, Foster-2015} to simulate the effects of unresolved binaries on our IVD measurements following a similar procedure as described in \citet{DaRio-2017}, \citet{Karnath-2019}, and \citetalias{Theissen-2022}. $235$ systems are generated across the binary fraction range from $0$ to $1$ with a step size of $0.25$, as is the total number of NIRSPEC and APOGEE sources used to generate the IVD.

The synthetic simulated RV distribution consists of three components: the systematics, the measurement uncertainty, and the binary offset. We briefly describe the steps to reproduce the three components below. First, a random intrinsic velocity dispersion is drawn from a uniform distribution between $1$ and $4.5~\mathrm{km}\,\mathrm{s}^{-1}$, which roughly covers a symmetric range on both ends lower and higher than the measured radial velocity dispersion of $2.87\pm0.15~\mathrm{km}\,\mathrm{s}^{-1}$. The systematics are the product of the intrinsic velocity dispersion and a standard normal distribution of length $235$. The simulated measurement uncertainty is generated by randomly sampling from the cumulative distribution function (CDF) constructed from the observed measurement error distribution, and multiplied by another standard normal distribution. Similarly, a mass distribution is generated by sampling from the CDF constructed by the interpolated mass distribution of the $235$ sources. An ensemble of stellar binary systems with uniform distribution of mass ratio and eccentricities is then simulated using this mass distribution with \texttt{velbin}, which in turn gives the radial velocity offset given the binary fraction.

The above process is repeated, and the first $10^5$ simulations are kept in which the standard deviation of the synthetic distribution lies within $2\sigma$ from the standard deviation of the observed distribution, where $\sigma$ is the uncertainty in the velocity dispersion in the radial component, or $0.15~\mathrm{km}\,\mathrm{s}^{-1}$ in our case. Note that radial velocity offsets with an absolute value greater than $7~\mathrm{km}\,\mathrm{s}^{-1}$ are truncated when fitting for the standard deviations to avoid the impact of extreme values.

Figure~\ref{fig:binary histogram} shows the simulated and observed velocity distributions in one of the simulations with binary fraction set to $50\%$. Figure~\ref{fig:binary violin} shows the distribution of the $10^5$ simulated IVDs that satisfy the aforementioned criterion under different imposed binary fractions. The blue violin plot illustrates the distribution of the simulated radial velocity dispersion. For comparison, IVDs in the radial, right ascension, and declination directions are shown in red dotted lines, green dash-dotted lines, and purple dashed lines, respectively. The crossing point between the measured radial velocity dispersion and the interpolation of the median of the simulated radial velocity dispersion at each binary fraction is at $62.54\%$. In other words, a binary fraction $\gtrsim~62.54\%$ is required to account for our measured higher velocity dispersion in the radial direction if the higher value is solely induced by binaries, which is unreasonably high compared to estimates in the literature. As can be seen from Figure~\ref{fig:binary violin}, the IVD of the radial component is higher than the other two directions, regardless of the simulated binary fraction. Therefore, the binarity alone is not sufficient to explain the larger values in the radial component. 


\subsubsection{Unresolved Binary Mass}
Unresolved close binaries are a potential source of systematic effects that could influence our results. Despite a lack of measurement of spectroscopic binaries in the region, we utilized multiplicity surveys collected in \citet{Offner-2023} and conducted a quantitative test on how the unresolved close binaries affect the correlation between relative velocity and stellar mass. Since Keck has a spatial resolution of $10$~au at the distance of the ONC, we adopted the close binary fraction (CBF) within $10$~au for brown dwarfs and main sequence stars to calculate the mass contribution from unresolved binaries. Historically, the distribution of the mass ratio $q$ is approximated by a power-law $f_q\propto q^\gamma$. We chose the values of $\gamma$ for binaries with a separation between $1$~au and $10$~au. The expectation of the mass of the hidden companion is therefore
\begin{equation}
    \mathrm{CBF}\times M\times\frac{\int_0^1 q\cdot f_q dq}{\int_0^1 f_q dq}~,
    \label{eq:binary mass}
\end{equation}
where the values of CBF and $q$ is determined by the observed primary mass according to \citet{Offner-2023}. Specifically, we utilized the values from the surveys in \citet{Winters-2019, Raghavan-2010}, and \citet{Tokovinin-2014} in order of increasing stellar mass. Note that $\gamma$ is assumed to be $0$ where the value is unavailable. The closest available values are utilized for sources situated in gaps of stellar mass ranges. For overlapping mass ranges, we adopt the values in the survey where the source is closer to the center of its mass range. Consequently, the adopted values are
\begin{equation}
\begin{aligned}
    &\mathrm{CBF}=16\%,&\gamma=0\quad&\mathrm{if}\quad0.075<M<0.15~,\\
    &\mathrm{CBF}=14\%,&\gamma=0.7\quad&\mathrm{if}\quad0.15<M<0.3~,\\
    &\mathrm{CBF}=15\%,&\gamma=0.1\quad&\mathrm{if}\quad0.3<M<0.675~,\\
    &\mathrm{CBF}=20\%,&\gamma=0.2\quad&\mathrm{if}\quad0.675<M<1.0875~,\\
    &\mathrm{CBF}=14\%,&\gamma=0.4\quad&\mathrm{if}\quad1.0875<M<1.5~.
\end{aligned}
\end{equation}
Substituting the above values back into Equation~\ref{eq:binary mass} for each source gives the expectation of the mass of the unresolved close binary.  As a result, the stellar mass increases by $8$--$14\%$.  Additionally, the negative correlation in Figure~\ref{fig:vrel-mass} persists after accounting for this excess in mass. The unresolved binaries only produce a marginally flatter linear fit, but still with a negative slope. For a quantitative comparison, the slopes in Figure~\ref{fig:vrel-mass} becomes $-0.75\pm0.18$, $-1.05\pm0.19$, $-0.71\pm0.19$, and $-0.72\pm0.14$ for MIST, BHAC15, Feiden, and Palla models respectively.  This is because the stellar mass on the higher-mass end is shifted further right in Figure~\ref{fig:vrel-mass}, while the change in mass of the lower-mass source is limited.  Therefore, the excess mass stretches the distribution of the data points horizontally, but the negative correlation persists.

\section{Discussion}
\label{sec:discussion}

\subsection{Kinematic Structure of the ONC and Star Formation Implications}

In this analysis, we have reconfirmed the supervirial nature of the central ONC.  This is primarily driven by the measurement of higher velocity dispersion in the radial dimension, which is not due to unresolved binaries.  We have also, for the first time, identified a tentative negative trend in relative velocity of stars as a function of mass, with lower mass stars having higher velocities than higher mass stars.  This has potentially strong implications for star formation.

The primary pathway of stellar formation across the full mass range in stellar clusters remains uncertain. \citet{Bonnell-2008} conducted hydrodynamical simulations to investigate low-mass star and brown dwarf formation in clusters. They argue that the filament-shaped infalling gas that is accreted onto a star cluster has high densities, allowing low-mass stars and brown dwarfs to form. However, the high velocity and tidal shear within the gas preclude those low-mass objects from accreting significantly from their surroundings any further. Therefore, one observable feature would be lower mass stars having higher velocities relative to their neighbors and vice versa. The baseline assumptions of this simulation are well-matched to the ONC: a young cluster residing within a filament of gas. Consequently, the ONC can serve as a perfect laboratory for this theory, which could provide keys to the origin of the initial mass function (IMF). 

Our measurements of a negative correlation between velocity and mass potentially indicate that the initial mass of forming stars may indeed depend on their initial kinematic states, supporting the gravitational fragmentation mechanism \citep[][]{Bonnell-2008}. Note that the negative correlation is more significant in the simulation in \citet{Bonnell-2008}, as the simulated cluster is still in its nascent age when inspected, only about $0.39$~Myr after the first stars formed. The negative trend in the ONC is already partially washed away by dynamical relaxation considering its age of $2$~Myr. 

Apart from \citet{Bonnell-2008}, the same trend between velocity and stellar mass is also detected in magnetohydrodynamical simulations of star cluster formation \citep[][]{Mathew-2021}. The velocities of the sink particles are found to be negatively correlated with their masses shortly after the birth of the stars. The disappearance of the correlation is observed as well over time due to dynamical evolution. We suggest that the ONC is undergoing a similar process of losing the currently observed trend between relative velocity and stellar mass.

Note that the RV measurements of the original APOGEE results were previously found to be negatively-correlated with $T_\mathrm{eff}$ for low temperature sources below $3400$~K \citep[][]{Cottaar-2014, Kounkel-2019}, which would introduce a bias in the velocity mass correlation. However, the reanalyzed APOGEE values adopted in this work are not affected by the bias, as we checked the consistency with the RVs in \citet{Kounkel-2019} after removing the systematics. The values conform well, with a median absolute difference of $0.46~\mathrm{km}\,\mathrm{s}^{-1}$.

There are interesting implications of our finding that velocity dispersion in the radial component is larger than that of the proper motion. We speculate that this can be attributed to the influence of the integral-shaped filament (ISF), a gas filament associated with the ONC \citep{Bally-1987}. The ISF is believed to experience periodic oscillations in radial and on-sky directions \citep[e.g.,][]{Stutz-2016, Stutz-2018, Matus-2023}. Protostars are ejected from the gas filaments during the oscillations, shutting off their accretion. Such a process of producing protostars is referred to as the slingshot mechanism. The ONC may be a star cluster that is radially ejected from the ISF towards us, which results in higher velocities and dispersion in the radial direction compared to the proper motion components.

The observed expansion of the ONC \citep{Kounkel-2022} and the super-virial velocity dispersion can be explained by the slingshot mechanism as well, according to \citet{Matus-2023}. After being ejected from the ISF, the decrease in gas within the ONC reduces the gravitational potential, resulting in the expansion and super-virial velocity dispersion. Alternatively, \citet{Kounkel-2022} argues that the expansion is driven by the unstable N-body interactions. Additional observations and tests are required to unveil the reasons for expansion.

\subsection{3D Spatial Kinematics: Parallax Simulation}
\label{subsec:parallax simulation}
\begin{figure*}[htb!]
    \centering
    \includegraphics[width=\textwidth]{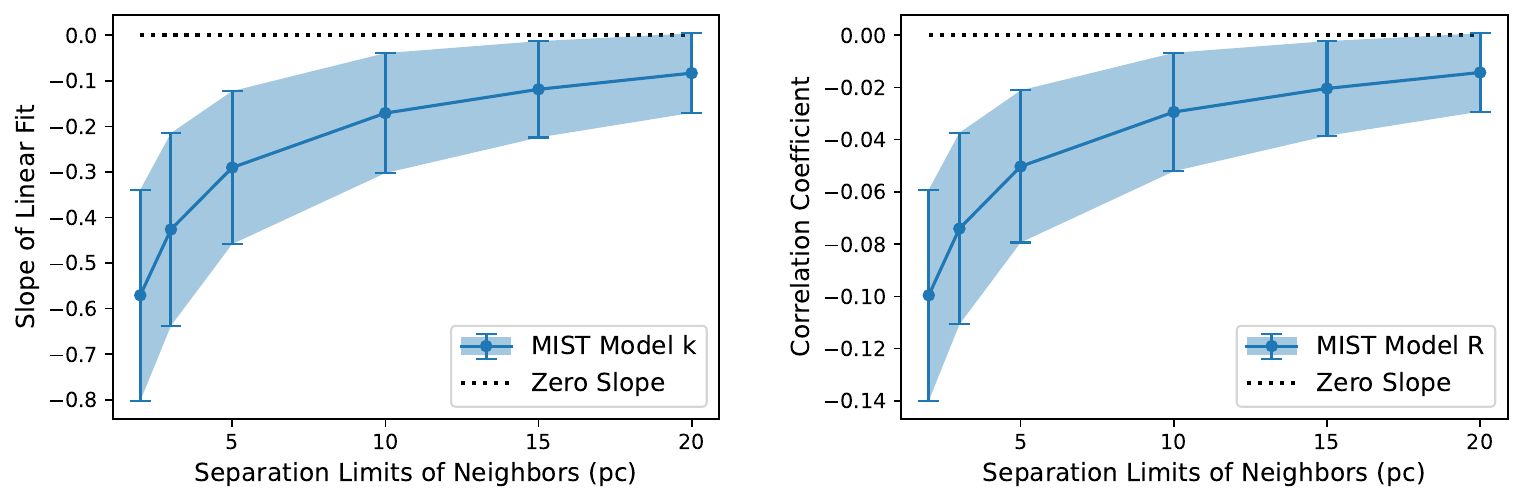}
    \caption{Parallax simulation results of the slope of linear fit $k$ and correlation coefficient $R$ between relative velocity and stellar mass under the MIST model as a function of separation limits within which sources are considered as neighbors. The blue errorbar and the shaded region show the value and associated uncertainty of the two parameters, while the dashed line shows the zero slope, or no correlation.}
    \label{fig:simulate k R}
\end{figure*}

Previous analysis was conducted under the assumption that all of the sources are located at the exact same distance of $389\pm3$~pc for reasons discussed in Section~\ref{subsec:3d velocities}. However, this is not the actual case, especially when we are considering the distances between neighboring sources. Seemingly adjacent sources on the plane of the sky might be distant from one another along our line of sight.

To investigate the impact of unknown distances to individual stars, we chose to simulate the parallax for all of our $235$ sources following the same distribution of the limited $100$ adopted \textit{Gaia} parallaxes. An inverse cumulative distribution function (CDF) is constructed from the $100$ parallax measurements. We then sample $235$ simulated parallaxes from the CDF and assign one value to each source. The simulation is repeated $1000$ times.

First, we present the simulation results on the correlation between relative velocity and stellar mass. The distance between sources is now updated to incorporate both the projection on the sky and the radial component. Figure~\ref{fig:simulate k R} shows the slope $k$ of the linear fit and correlation coefficient $R$ between relative velocity and stellar mass as a function of the separation limit within which we consider sources as neighbors. The blue errorbar and the shaded region represent the mean and standard deviation of the $1000$ simulated results of the slope $k$ and correlation coefficient $R$. As can be seen, the same increasing trend as the separation limit increases persists for both parameters, consistent with Figure~\ref{fig:slope_vs_mass}. Therefore, the conclusion remains based on the $1000$ parallax simulations: the negative correlation between relative velocity and stellar mass exists and becomes increasingly obvious when we consider the velocity relative to the more immediate neighbors of each source.

\begin{figure}[htb!]
    \centering
    \includegraphics[width=\columnwidth]{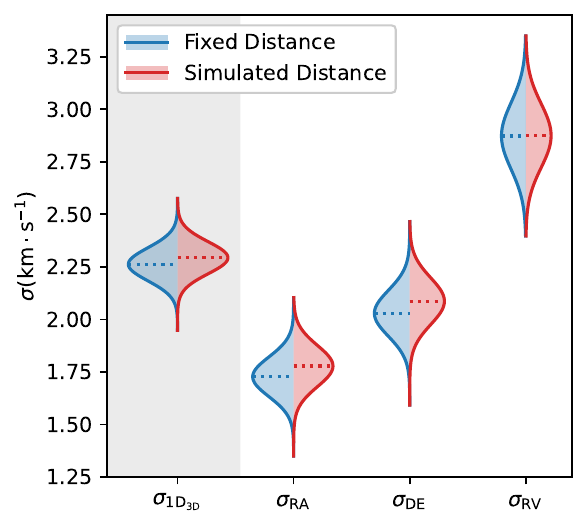}
    \caption{Comparison of velocity dispersions between the fixed distance scenario and the parallax simulation. The blue profile on the left of each column is a normal distribution of the velocity dispersion when every source is assumed at a fixed distance of $389\pm3$~pc, with the mean and the standard deviation specified as in Equation~\ref{eq:vdisp measurements}. The red profile on the right of each column shows the normalized summation of the $1000$ normal distributions of the velocity dispersions. We then fit a single normal distribution over the summed distribution, and the dashed lines show the mean of the normal distributions. The $\sigma_\mathrm{1D_{3D}}$ is shaded with a light gray background to visually distinguish from its components in all three directions.}
    \label{fig:vdisp simulate}
\end{figure}

Second, we report the velocity dispersion measurements in the parallax simulation. The same procedure of measuring the velocity dispersion is conducted in each simulation as in Section~\ref{subsec:virial}. Each of the $1000$ simulations produces velocity dispersions and their associated uncertainties. To determine the final uncertainty of the simulation, we first add $1000$ normal distributions centering at the simulated values with the standard deviation being the uncertainties of each simulation and normalize it with a factor of $1/1000$. Then we fit a single normal distribution to the summation, and use the mean and standard deviation as the simulated values and uncertainties. The summation is justified because we are intrinsically assuming a normal posterior distribution of the parameters when we use a normal distribution likelihood function in the MCMC ensemble sampler. The simulated velocity dispersions are listed below:
\begin{align} \label{eq:vdisp simulation}
    \overline{\sigma}_\mathrm{RA} = 1.78\pm0.10~\mathrm{km}\,\mathrm{s}^{-1}, \nonumber \\
    \overline{\sigma}_\mathrm{DEC} = 2.09\pm0.11~\mathrm{km}\,\mathrm{s}^{-1}, \nonumber \\
    \overline{\sigma}_\mathrm{RV} = 2.87\pm0.15~\mathrm{km}\,\mathrm{s}^{-1}, \nonumber \\
    \overline{\sigma}_\mathrm{1D_{3D}} = 2.29\pm0.08~\mathrm{km}\,\mathrm{s}^{-1}.
\end{align}
A comparison of the velocity dispersions between the fixed and simulated distance scenarios is shown in Figure~\ref{fig:vdisp simulate}. As can be seen, the 1D velocity dispersion with simulated parallax is only slightly boosted by $0.03~\mathrm{km}\,\mathrm{s}^{-1}$ from the case of fixed distance, even smaller than the uncertainty. The change is mostly caused by a minute increase in the RA and DEC components, with the RV component remaining exactly the same. This proves that the projection effect cannot account for the higher $\sigma_\mathrm{RV}$ than the other $2$ directions. 

Additionally, the simulated virial ratio is $q=0.88$, only slightly different from the previous result of $0.85$. In summary, the projection on the plane of the sky does not have a significant effect on the velocity dispersion results.

\subsection{Mass Segregation and Energy Equipartition}
\label{subsec:segregation}
\begin{figure*}[htb!]
    \centering
    \includegraphics[width=\columnwidth]{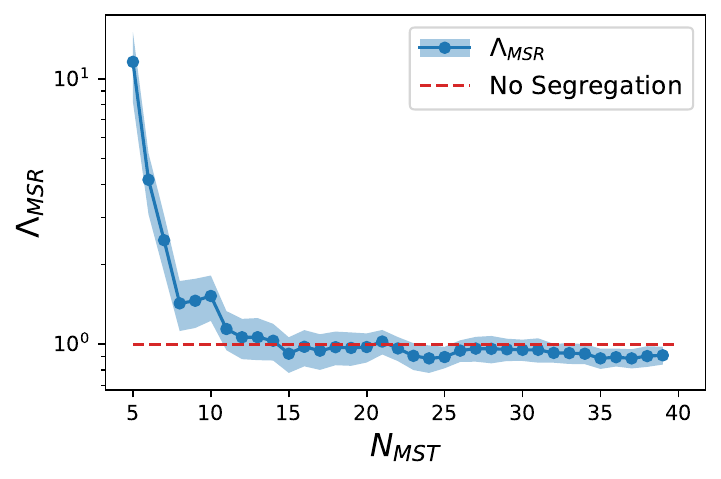}
    \includegraphics[width=\columnwidth]{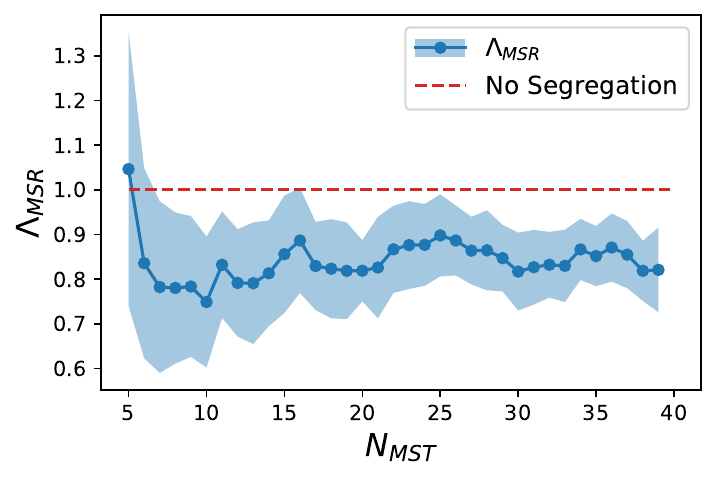}
    \caption{Mass segregation ratio $\Lambda_\mathrm{MSR}$ under MIST model as a function of the number of sources chosen to construct the minimum spanning tree $N_\mathrm{MST}$. \textit{Left}: Including the Trapezium stars. \textit{Right}: Excluding the Trapezium star. Values greater than $1$ indicate mass segregation for the corresponding number of most massive stars, and inverse mass segregation for $\Lambda_\mathrm{MSR}$ less than $1$. The dividing line is shown in a dashed red line.}
    \label{fig:segregation}
\end{figure*}

Mass segregation refers to the non-random distribution of stars in stellar systems, where more massive stars tend to concentrate toward the center while lower-mass stars are more dispersed in the outer regions. This occurs due to gravitational interactions and the differential response of stars of different masses to these interactions. Understanding mass segregation and its origin is essential for studying the evolution and formation of stellar systems \citep[e.g.,][]{Fregeau-2002, Baumgardt-2008}.

With stellar masses derived in Section~\ref{subsec:mass}, the mass segregation in the central ONC can be quantified. Here we adopt the mass segregation ratio (MSR), $\Lambda_\mathrm{MSR}$, defined in \citet{Allison-2009}. This parameter uses the minimum spanning tree to measure how clustered the most massive stars in a cluster are. If the most massive stars tend to stay closer to each other compared to a random group of stars of the same amount, the ratio is larger than $1$, indicating mass segregation. Conversely, if the most massive stars are further away from each other than a random combination of stars, the ratio is less than $1$, or inversely mass segregated.

Figure~\ref{fig:segregation} shows the MSR under the MIST model as a function of $N_\mathrm{MST}$. Figure~\ref{fig:segregation}(a) shows our target sources combined with the high-mass Trapezium stars, while in Figure~\ref{fig:segregation}(b) the Trapezium stars are excluded to unveil the mass segregation for other stars. The dividing line of $\Lambda_{MSR}=1$ is plotted as a red dashed line in both figures. It is evident from Figure~\ref{fig:segregation}(a) that the $5$ most massive stars are very mass segregated, which is expected as the Trapezium stars are clearly spatially clustered in the ONC center.  We utilized the masses of the Trapezium stars from the literature (\citealt{Weigelt-1999, Close-2012} for $\theta^1$ Orionis A, \citealt{Vitrichenko-2006} for $\theta^1$ Orionis B, \cite{Balega-2014} for $\theta^1$ Orionis C, \citealt{Allen-2017} for $\theta^1$ Orionis D, and \citealt{Morales-Calederon-2012} for $\theta^1$ Orionis E). \citet{Hillenbrand-1998} argues that the high-mass stars are likely prone to form in the ONC center. Surprisingly, Figure~\ref{fig:segregation}(b) shows that the stars apart from the few Trapezium stars display a feature of inverse mass segregation, i.e., massive stars tend to be on the outskirts. This result may not be a true feature and could instead be a result of our sample selection. As mentioned previously, our goal in the NIRSPAO target selection was to obtain a good sampling of low-mass objects in the cluster core. The APOGEE sources are not preferentially sampled at low masses, tending to higher masses because of higher mass sensitivity limits at the $H$-band.  The APOGEE sources are also generally on the periphery of the region we are analyzing. Therefore, we may have missed some sources on the lower-mass end in the outskirts, which results in a bias in this result.  More complete spectra in the cluster core and outskirts are needed to assess whether this trend is real or a sample artifact.

While the most massive Trapezium stars in the cluster exhibit a close distribution around the cluster center, it is important to recognize that mass segregation does not necessarily indicate the occurrence of energy equipartition, as N-body simulation suggests \citep[][]{Parker-2016}. Instead, the kinetic energy decreases at the same rate for low and high-mass stars according to the simulation. This is consistent with our observation in that the stars with higher masses still have a slightly higher velocity dispersion, as can be seen from the left panel in Figure~\ref{fig:vdisp vs mass}.

\section{Conclusions}
\label{sec:conclusion}
In this work, we present the 3D kinematic analysis of $235$ sources in the central region of the ONC, including $80$ sources observed by the Keck NIRSPAO and $167$ reanalyzed APOGEE sources, with $17$ common sources between the two surveys. High-precision radial velocities and effective temperatures, along with other stellar parameters, are retrieved from spectral analysis. With the help of previous proper motion measurements from HST and multiple stellar evolutionary models, we construct a 3D kinematic map of the region with interpolated stellar mass. Listed below are the main takeaways for this work.

\begin{enumerate}
    \item We model stellar parameters including $T_\mathrm{eff}$ and RV using high-resolution spectra from NIRSPAO ($R\sim25000$ or $35000$). Stellar masses are then derived from $T_\mathrm{eff}$ and four different stellar evolutionary models assuming a stellar age of $2\pm1$~Myr.

    \item The velocity dispersion in the right ascension, declination, radial direction are $\sigma_\mathrm{RA} = 1.73 \pm 0.09~\mathrm{km}\,\mathrm{s}^{-1},~\sigma_\mathrm{DEC} = 2.03 \pm 0.11~\mathrm{km}\,\mathrm{s}^{-1},~\sigma_\mathrm{RV} = 2.87 \pm 0.16~\mathrm{km}\,\mathrm{s}^{-1}$. The 1D velocity dispersion is $\sigma_\mathrm{1D_{3D}} = 2.26 \pm 0.08~\mathrm{km}\,\mathrm{s}^{-1}$, exceeding the requirement of virial equilibrium of $1.73~\mathrm{km}\,\mathrm{s}^{-1}$ \citep[][]{DaRio-2014}. Therefore, the ONC is not fully virialized yet, with a virial ratio of $0.85$, consistent with \citet{DaRio-2014}.

    \item A negative correlation between the velocity relative to the neighbors of each source and the stellar mass is identified using four different stellar evolutionary models, consistent with gravitational filament fragmentation simulation results. This suggests that during the star formation processes within infalling gas filaments, the high velocities of the fast-moving primordial stars preclude them from accreting more material from their surroundings.

    \item Neither the velocity dispersion nor the negative correlation is affected by the projection effect by assuming the same distance of $389\pm3$~pc for all sources. We conducted $1000$ simulations in which each source is assigned a simulated distance according to the distribution of the $100$ adopted \textit{Gaia} parallax measurements within our sample. The simulated IVD is consistent with the projected scenario, with $\sigma_\mathrm{RA}$, $\sigma_\mathrm{DEC}$, and $\sigma_\mathrm{1D_{3D}}$ slightly increased by less than $1\sigma$. The negative correlation still holds.
    
    \item There is a systematic discrepancy in the effective temperature between NIRSPAO and APOGEE results. The difference could be attributed to the different passbands used in the two instruments, $K$ and $H$ bands, respectively. The negative correlation between relative velocity and stellar mass still exists after accounting for the discrepancy by offsetting the NIRSPAO $T_\mathrm{eff}$ by the weighted-averaged difference. Furthermore, the negative correlation becomes more significant for more localized relative velocity.

    \item A clockwise rotational preference in proper motions is identified in the region.

    \item The sources are found to be inversely mass segregated, or massive stars being more scattered on the outskirts, if the Trapezium stars are excluded. This may stem from a selection bias as we are focusing on low-mass stars closer to the ONC center.

    \item We report $4$ binary candidate systems by observing the change in radial velocity, Parenago 1837, V* V1337 Ori, V*1279 Ori, and Brun 590. We were able to infer the properties of the companion with a mass most likely of $\sim0.03$--$0.3~M_\odot$ and a semi-major axis most likely of less than $2$~au.

    \item We simulate the effect of binaries have on the IVD in the radial component and conclude that binarity alone is insufficient to explain the higher value in IVD in the radial direction compared to the proper motion directions. Possible explanations are the interaction with the ISF, such as the slingshot mechanism, which states that the ONC is ejected from the ISF due to the oscillation of the gas filament \citep[][]{Stutz-2016, Stutz-2018, Matus-2023}.

    \item We calculate the expectation of the excess in mass contributed by unresolved close binaries within $10$~au. The expectation of the increase in mass is $8$--$14\%$. The negative correlation between relative velocity and stellar mass persists, albeit with a marginally flatter linear fit, after accounting for the contribution from unresolved binary mass. 

\end{enumerate}

In the future, we plan to construct kinematic maps in an extended area beyond our current $4\arcmin$ radius. This may place a tighter constraint on the current kinematic state of the larger cluster population, probing deeper into its formation history and implications. In addition, our observations are ideal first epochs for unresolved binary star searches, which will provide a true estimate of the very tight binary fraction in this region.

\section{Acknowledgments}
We would like to thank an anonymous referee whose suggestions greatly improved this manuscript.  L.W. thanks the LSSTC Data Science Fellowship Program, which is funded by LSSTC, NSF Cybertraining Grant \#1829740, the Brinson Foundation, and the Moore Foundation; his participation in the program has benefited this work. Portions of this work were funded by NSF grant AST-1714816.

The data presented herein were obtained at the W. M. Keck Observatory, which is operated as a scientific partnership among the California Institute of Technology, the University of California, and the National Aeronautics and Space Administration. The Observatory was made possible by the generous financial support of the W. M. Keck Foundation. The authors wish to recognize and acknowledge the very significant cultural role and reverence that the summit of Maunakea has always had within the indigenous Hawaiian community. We are most fortunate to have the opportunity to conduct observations from this mountain.  

The authors thank support astronomers Carlos Alvarez, Greg Doppmann, Jim Lyke, Luca Rizzi, Hien Tran, and telescope operators Joel Aycock, Tony Connors, Heather Hershley, Julie Renaud-Kim, Gary Puniwai, Arina Rostopchina, Terry Stickel, and Cynthia Wilburn for their help in obtaining the observations.

Portions of this work were conducted at the University of California, San Diego, which was built on the unceded territory of the Kumeyaay Nation, whose people continue to maintain their political sovereignty and cultural traditions as vital members of the San Diego community.

This work has made use of data from the European Space Agency (ESA) mission
{\it Gaia} (\url{https://www.cosmos.esa.int/gaia}), processed by the {\it Gaia}
Data Processing and Analysis Consortium (DPAC,
\url{https://www.cosmos.esa.int/web/gaia/dpac/consortium}). Funding for the DPAC
has been provided by national institutions, in particular the institutions
participating in the {\it Gaia} Multilateral Agreement.

Funding for the Sloan Digital Sky Survey IV has been provided by the Alfred P. Sloan Foundation, the U.S. Department of Energy Office of Science, and the Participating Institutions. SDSS acknowledges support and resources from the Center for High-Performance Computing at the University of Utah. The SDSS website is www.sdss4.org.

SDSS is managed by the Astrophysical Research Consortium for the Participating Institutions of the SDSS Collaboration including the Brazilian Participation Group, the Carnegie Institution for Science, Carnegie Mellon University, Center for Astrophysics | Harvard \& Smithsonian (CfA), the Chilean Participation Group, the French Participation Group, Instituto de Astrofísica de Canarias, The Johns Hopkins University, Kavli Institute for the Physics and Mathematics of the Universe (IPMU) / University of Tokyo, the Korean Participation Group, Lawrence Berkeley National Laboratory, Leibniz Institut für Astrophysik Potsdam (AIP), Max-Planck-Institut für Astronomie (MPIA Heidelberg), Max-Planck-Institut für Astrophysik (MPA Garching), Max-Planck-Institut für Extraterrestrische Physik (MPE), National Astronomical Observatories of China, New Mexico State University, New York University, University of Notre Dame, Observatório Nacional / MCTI, The Ohio State University, Pennsylvania State University, Shanghai Astronomical Observatory, United Kingdom Participation Group, Universidad Nacional Autónoma de México, University of Arizona, University of Colorado Boulder, University of Oxford, University of Portsmouth, University of Utah, University of Virginia, University of Washington, University of Wisconsin, Vanderbilt University, and Yale University.


%

\newpage
\facilities{HST(ACS/WFPC2/WFC3IR), SDSS(APOGEE), Keck}


\software{
    \texttt{astropy} \citep{Astropy-2013, Astropy-2018, Astropy-2022},
    \texttt{astroquery} \citep{astroquery},
    \texttt{smart} \citep{Hsu-2021-code, Hsu-2021-paper},
    \texttt{emcee} \citep{Foreman-Mackey-2013},
    \texttt{velbin}  \citep{Cottaar-2014, Foster-2015},
    \texttt{TheJoker} \citep{thejoker}
}




\bibliography{main}{}
\bibliographystyle{aasjournal}


\end{CJK*}
\end{document}